\documentclass[amsmath,amssymb,aps,prl,twocolumn]{revtex4-1}    
\usepackage[pdftex]{graphicx}		

\usepackage{txfonts}
\usepackage{bm}
\usepackage{array}
\usepackage{amsmath}
\usepackage{amssymb}
\usepackage{color}
\usepackage[normalem]{ulem}	

\newcommand{\kCN}{$\kappa$-CN}
\newcommand{\kCNFull}{$\kappa$-(BEDT-TTF)$_2$Cu$_2$(CN)$_3$}
\newcommand{\kClFull}{$\kappa$-(BEDT-TTF)$_2$Cu[N(CN)$_2$]Cl}
\newcommand{\kCl}{$\kappa$-Cu-Cl}
\newcommand{\kHgBrFull}{$\kappa$-(BEDT-TTF)$_2$Hg(SCN)$_2$Br}
\newcommand{\kHgBr}{$\kappa$-Hg-Br}
\newcommand{\kHgClFull}{$\kappa$-(BEDT-TTF)$_2$Hg(SCN)$_2$Cl}
\newcommand{\kHgCl}{$\kappa$-Hg-Cl}
\newcommand{\dmit}{EtMe$_3$Sb[Pd(dmit)$_2$]$_2$}

\newcommand{\TMI}{T_{\textrm{MI}}}
\newcommand{\TCW}{\Theta_{\textrm{CW}}}

\begin{document}

	\title{Ferromagnetism out of charge fluctuation of strongly correlated electrons in \kHgBrFull}
	
	\author{Minoru Yamashita$^1$}
	\email[]{my@issp.u-tokyo.ac.jp}
	\author{Shiori Sugiura$^{2}$}
	\altaffiliation[Present address: ]{Institute for Materials Research, Tohoku University, Katahira 2-1-1, Aoba-ku, Sendai, Miyagi 980-8577, Japan}
	\author{Akira Ueda$^{1}$}
	\altaffiliation[Present address: ]{Department of Chemistry, Faculty of Advanced Science and Technology, Kumamoto University, 2-39-1 Kurokami, Chuo-ku, Kumamoto 860-8555, Japan}
	\author{Shun Dekura$^{1}$}
	\author{Taichi Terashima$^{2}$}
	\author{Shinya Uji$^{2}$}	
	\author{Yoshiya Sunairi$^{1}$}
	\author{Hatsumi Mori$^{1}$}
	\author{Elena I. Zhilyaeva$^{3}$}
	\author{Svetlana A. Torunova$^{3}$}
	\author{Rimma N. Lyubovskaya$^{3}$}\thanks{Deceased.}
	\author{Natalia Drichko$^{1,4}$}
	\email[]{drichko@jhu.edu}
	\author{Chisa Hotta$^{5}$}
	\email[]{chisa@phys.c.u-tokyo.ac.jp}

	\affiliation{
		$^1$The Institute for Solid State Physics, The University of Tokyo, Kashiwa, Chiba 277-8581, Japan\\
		$^2$National Institute for Materials Science, Tsukuba, Ibaraki 305-0003, Japan\\
		$^3$Institute of Problems of Chemical Physics RAS, Chernogolovka, Moscow region, 142432 Russia\\
		$^4$The Institute for Quantum Matter and the Department of Physics and Astronomy, The Johns Hopkins University, Baltimore, MD 21218, USA\\ 
		$^5$Department of Basic Science, University of Tokyo, Meguro-ku, Tokyo 153-8902, Japan
	}
	
	\date{\today}
	
	\begin{abstract}
	We perform magnetic susceptibility and magnetic torque measurements on the organic {\kHgBrFull}, which is recently suggested to host an exotic quantum dipole-liquid in its low-temperature insulating phase. Below the metal-insulator transition temperature, the magnetic susceptibility follows a Curie-Weiss law with a positive Curie-Weiss temperature, and a particular $M\propto \sqrt{H}$ curve is observed. The emergent ferromagnetically interacting spins amount to about 1/6 of the full spin moment of localized charges. Taking account of the possible inhomogeneous quasi-charge-order that forms a dipole-liquid, we construct a model of antiferromagnetically interacting spin chains in two adjacent charge-ordered domains, which are coupled via fluctuating charges on a Mott-dimer at the boundary. 
We find that the charge fluctuations can draw a weak ferromagnetic moment out of the spin singlet domains. 
	\end{abstract}

       \maketitle
       
\section{Introduction}

       Typical phase transitions in condensed matter accompany either universal critical singularities or 
       the competitions between two different orderings. 
       The former is easily converted to the latter when additional degrees of freedom become relevant. 
       In reality, there often appear intermediate situations where the interplay of several degrees of freedom affects the nature of the growth of correlations and low-lying excitations. 
       In such cases, the phase transitions at low temperatures can be easily masked, and the order parameters suffer intrinsic inhomogeneities. 
       Historical examples are the dynamically disordered charge stripes in high-$T_{\rm c}$ superconducting 
       cuprates~\cite{Kivelson1998,Kivelson2003}, 
       and the orbital-disorders that trigger the colossal magnetoresistance in manganites~\cite{SalamonJaime2001}. 
       The complexity of dealing with multiple correlated degrees of freedom such as charge, spin, 
       orbital, and lattice often makes it difficult to pin down their dominant mechanism. 

\begin{figure}[!tbh]
	\centering
	\includegraphics[width=9cm]{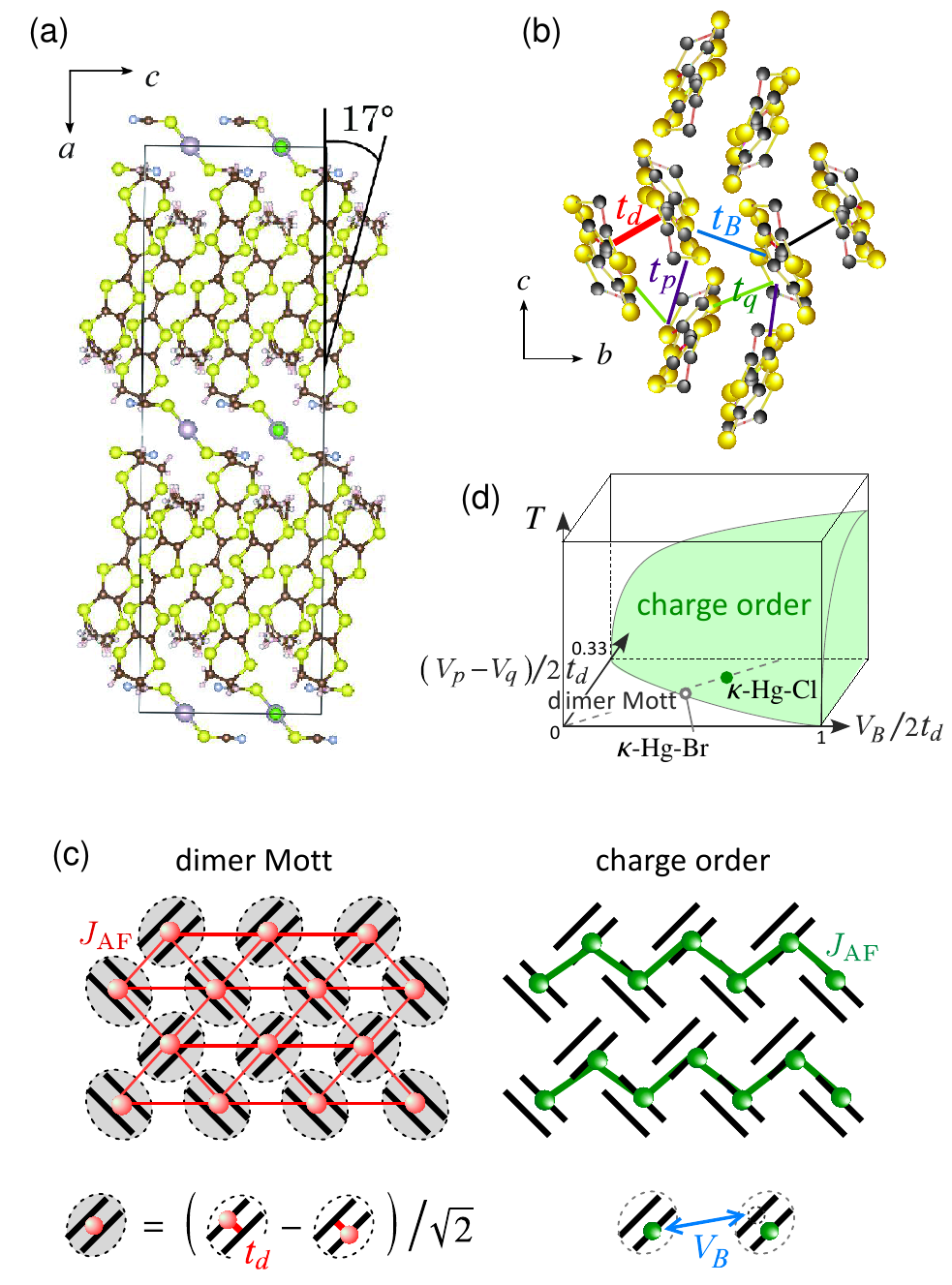}	
	\caption{
		(a) Crystal structure of {\kHgBr} viewed along the $b$-axis. 
		The long molecular axis of BEDT-TTF is tilted by $\sim$17$^\circ$ from the $a$-axis. 
		(b) BEDT-TTF molecules in the $b$-$c$ plane and the transfer integrals estimated as 
		$(t_d,t_B,t_p,t_q)=(126,83,60,40)$~meV in Ref.~\cite{ValentiJeschkePrv, Gati2018}. 			
		(c) Schematic illustration of the dimer Mott insulator and charge order. 
		Charges on Mott-dimers (red circles) carry spin-1/2, 
		and the exchange interactions between them 
		$J_{\rm AF}\propto t_B^2$ (vertical) and $J_{\rm AF}\propto(t_p-t_q)^2$ (diagonal directions)
		form an antiferromagnetic triangular lattice (red lines). 
		In the charge ordered state, the charges localized on one side of the dimer (green circles) 
		form an antiferromagnetic quasi-one-dimensional spin-1/2 chain of $J_{\it AF}\propto t_q^2$ (green line). 
		Inset: the Mott-dimer state is the linear combination of charge located on the left/right molecule, 
		supported by large $t_d$. The charge order keeps the charge on one side of the dimerized pair 
		to avoid the inter-site (inter-dimer) Coulomb interaction $V_{ij}=V_B,V_p,V_q$ (indices follow those of $t_{ij}$). 
		(d) Phase diagram (schematic)~\cite{Hotta-Harada} of the present system 
		for low energy effective model of charges proposed in Ref.~[\onlinecite{Hotta2012}], 
		where $(V_p-V_q)$ and $V_B$ account for the Coulomb interactions between charges on different dimers in diagonal and vertical directions, respectively. When $V_{ij}/2t_d\gtrsim 0.5$--1, the charge order is realized. 
		According to the first principles-based evaluation~\cite{Jacko2020}, 
		$((V_p-V_q)/2t_d, V_B/2t_d)\sim (0.25,0.48)$ for {\kHgCl} and $(0.19,0.42)$ for {\kHgBr}. 
	}
	\label{fig1}
\end{figure}

       Organic $\kappa$-(BEDT-TTF)$_2 X$ materials (Fig.\,\ref{fig1}(a) and (b)) become an ideal playground 
       to study such an issue in a simpler setup. 
       These materials form quarter-filled two-dimensional strongly correlated electronic systems, 
       where molecular dimer (BEDT-TTF)$_2$ connected by a large transfer integral $t_d$ (see Fig.\,\ref{fig1}(b))
       serves as a lattice site of a Mott insulator by hosting one charge per dimer. 
       In these Mott insulators, the spin-1/2 interact antiferromagnetically as shown Fig.\,\ref{fig1}(c) 
       and a quantum spin liquid phase is observed in {\kCNFull} (abbreviated as {\kCN}) 
       as well as a typical antiferromagnetism in {\kClFull} ({\kCl})~\cite{Kanoda2006,KanodaKato2011}. 
       If the degree of dimerization, namely the ratio of $t_d$ to other inter-dimer transfer integrals ($t_B$, $t_p$, and $t_q$), 
       is weakened, the charges would no longer stay on the dimer-orbital, 
       but rather localize to one side of the dimerized molecules 
       to gain the inter-molecular Coulomb interaction $V_{ij}$, resulting in a charge-ordered phase. 
       A charge degree of freedom enclosed in the dimer is interpreted as quantum electric dipole~\cite{Hotta2010,Majed2010}, 
       which is detected by anomalous frequency-dependence of dielectricity in many materials 
       including {\kCN}~\cite{Majed2010}, {\kCl}~\cite{Lunkenheimer2012}, and {\dmit}~\cite{Majed2013, Lazic2018}. 
       In that context, the dimer Mott and charge-ordered phases are interpreted as para 
       and ferroelectricity~\cite{Hotta2010,NakaIshihara2010}, 
       separated by a typical Ising type second-order phase transition (see the phase diagram in Fig.\,\ref{fig1}(d)). 
       However, when the universal criticality of dipoles couples to magnetism or lattice degrees of freedom, 
       this transition can be masked and some inhomogeneous phases may emerge~\cite{Itou2017}. 
       Indeed, the subtleties of the transition are recently disclosed by the fresh members of 
       this family, {\kHgBrFull} ({\kHgBr}) and {\kHgClFull} ({\kHgCl}), 
       which have a relatively weak dimerization~\cite{Gati2018} and fill the empty region of materials parameter space. 
       In contrast to a simple Mott insulator, which shows a crossover from the high-temperature metallic regime~\cite{Kanoda2006}, 
       these compounds show an abrupt increase of resistivity at the metal-insulator (MI) transition~\cite{Ivek2017}. 
       Raman spectroscopy reveals a distinct charge order in {\kHgCl} in the temperature range 15--30\,K~\cite{Hassan2020} 
       whereas {\kHgBr} does not show any sign of regular charge ordering down to lowest temperature~\cite{Hassan2018}.
       The absence of magnetic order in {\kHgBr} is also shown by the specific heat measurements down to 100\,mK~\cite{Hassan2018}.
       A picture of ``quantum dipole liquid'' is provided as an interpretation to the latter intriguing phase~\cite{Hassan2018}, 
       possibly consisting of dynamical charge-ordered domains enclosing electric dipole moments maximally amounting to $0.1e$ per dimer.  


       We report the experimental evidence of 
       intrinsic ferromagnetic exchange interactions emerging in the clean bulk crystal of {\kHgBr}, 
       indicated by a positive Curie-Weiss temperature of $\TCW = 16$~K, where about 1/6 of the full spins of localized charges contribute. 
       We find that the $M$--$H$ curve at low temperature follows $M\propto \sqrt{H}$, 
       showing a very rapid onset with small field. 
       Although the square-root {\it onset} of the $M$--$H$ curve is well-known for 
       a {\it gapped} quasi-one-dimensional quantum magnet near the critical field~\cite{Affleck1991,ChitraGiamarchi1997}, 
       it is qualitatively different from the present {\it gapless} $M\propto \sqrt{H}$ that continues up to a large field. 
       It does not resemble any of the $M$--$H$ profile of the magnetism known so far 
       such as the $H$-linear antiferromagnetic magnetization or the paramagnetic Brillouin curve. 
       Such robust ferromagnetic Curie-Weiss law just above the antiferromagnetic singlet ground state can be scarcely found in nature, 
       except for those originating from magnetic impurities or a spin glass, both of which are excluded in the present case 
        by the lack of remnant filed or hysteresis in magnetization.
	Since no existing theory on {\it bulk} magnetism both for the localized spins or itinerant electrons can be applied, 
	we construct a synergetic quantum-spin model that includes the effect of charge fluctuation. 
        The starting point is the low temperature inhomogeneous state of charges that appear by masking 
        the phase transition in Fig.~\ref{fig1}(d). 
        We take account of already existing idea of a short charge correlation length and the robustly remaining charge fluctuation 
        at the simplest level~\cite{Hotta-Harada,Hotta2012,Hassan2018}.
	The model represents spins on two charge-ordered domains which couple to dimer-spins carried by fluctuating charges at the domain boundary, and successfully shows how ferromagnetic behavior can originate from the charge fluctuation.
	The theory thus explains the properties disclosed by the magnetic susceptibility and torque measurements.


\begin{figure}[!tbh]
	\centering
	\includegraphics[width=8cm]{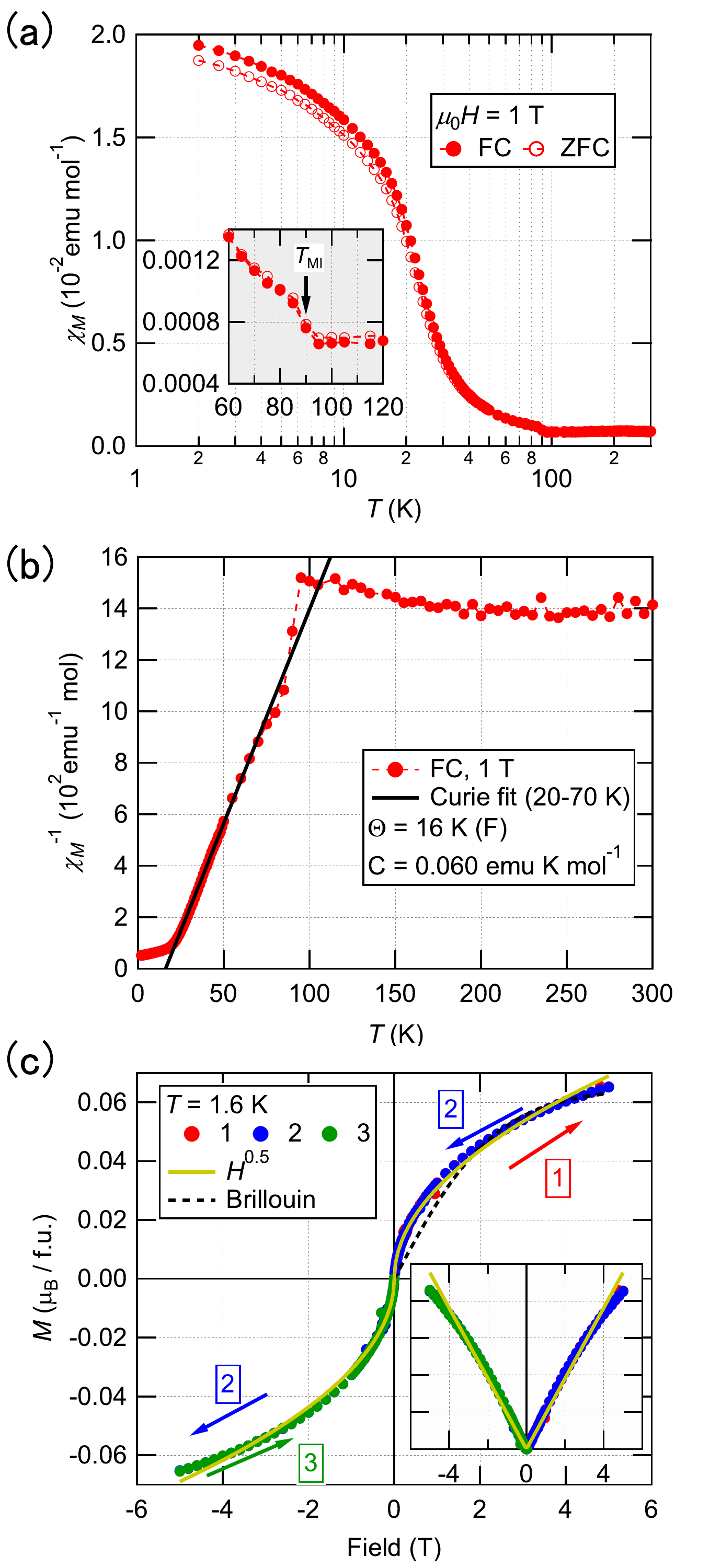}
	\caption{
		(a) Temperature dependence of the magnetic susceptibility ($\chi_M(T)$) at 1~T. Data obtained by field cool (FC) and zero-field cool (ZFC) are shown by filled and open symbols, respectively. The inset shows an enlarged view of $\chi_M(T)$ near $\TMI = 90$~K.
		(b) Temperature dependence of the inverse susceptibility. 
		The solid line shows a linear fit for 20--70 K.
		(c) Field dependence of the magnetization $M$ at 1.6 K, 
		measured in the order of (1) 0 to 5 T, (2) +5 T to -5 T, and (3) -5 T to 0 T. 
		The solid and the dashed lines show a fit to $M \propto \sqrt{H}$ and the Brillouin function at 1.6\,K, respectively. The inset shows the field dependence of $M^2$ of the same data. 
	}
	\label{SQUID}
\end{figure}

\section{Results}
\subsection{Magnetization measurements}

	Figures\,\ref{SQUID}(a) and (b) show the temperature dependence of the magnetic susceptibility $\chi_M(T)$ $(=M/H)$ of {\kHgBr}.
	The MI transition is observed at $\TMI\sim 90$ K (the inset of Fig.\,\ref{SQUID}(a)). 
        Above this temperature, $\chi_M(T)$ shows a Pauli-paramagnetic behavior, 
	while below $\TMI$ it starts to increase abruptly on lowering the temperature. 
	A Curie-Weiss fit for 20--70 K (the solid line in Fig.~\ref{SQUID}(b)) gives a \textit{positive} $\TCW \sim 16$~K with the Curie constant $C=0.060$~emu~K~mol$^{-1}$. 
	This positive $\TCW$ provides strong evidence of a ferromagnetic interaction between spins.
	The $C$ value shows that $\sim$1/6 of the total spins contribute to the Curie-Weiss paramagnetism with the ferromagnetic interaction. 
	The 1/6-concentration is intrinsic to the ferromagnetic behavior, 
        as $\chi_M(T)$ does not depend on the measured field strength below 5\,T in this temperature range (see Fig.\,S2 in  Supplemental Material (SM)~\cite{SM}). 
	Similar $\chi_M(T)$ was observed previously~\cite{Hemmida2018}. However, our data shows the ferromagnetic $\TCW$ more clearly in a wider temperature range (see Section A in SM~\cite{SM} for a comparison). 
	To the best of our knowledge, this compound is the first to show such ferromagnetic behavior in a family of organic Mott insulators $\kappa$-(BEDT-TTF)$_2 X$ and $X$[Pd(dmit)$_2$]$_2$. 
	If one interprets this $\chi$ as the one from the ferromagnetic Heisenberg chain\,\cite{Souletie2005}, 
	the effective ferromagnetic coupling constant is evaluated as $J_{F} \sim \TCW/\Theta = 53$\,K, with $\Theta=0.3036$.


\begin{figure}[t!bh]
	\centering
	\includegraphics[width=8cm]{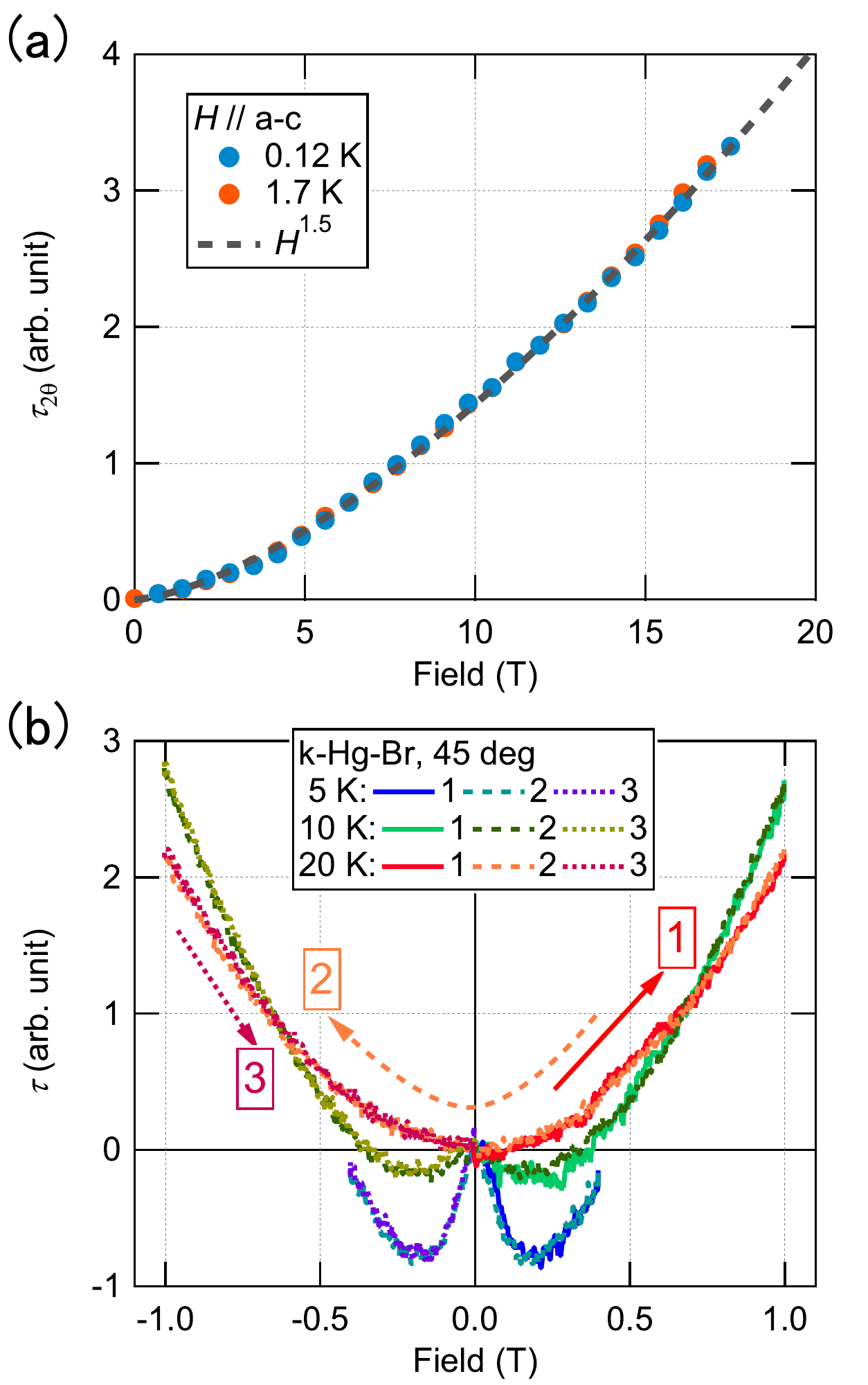}
	\caption{
		(a) Field dependence of the magnetic torque up to 17.5~T at 0.12 and 1.7~K. The amplitude of the $2 \theta$ component of the magnetic torque curve obtained by rotating magnetic field in the $a$-$c$ plane is plotted. 
		The dashed line shows a fit of $H^{3/2}$ field dependence, 
		indicating $M \propto \sqrt{H}$ up to 17.5\,T.
		(b) Low-field data of the magnetic torque at 5, 10 and 20~K. 
		The field angle was fixed at 45~degree from the $c$-axis to the $a$-axis. 
		Field sweep is given as (1) $0 \to +H_{max}$, (2) $+H_{max} \to -H_{max}$, and (3) $-H_{max} \to 0$. 
	}
	\label{torque_Hdep}
\end{figure}

The particular ferromagnetic behavior is also found in the field dependence of the magnetization $M$. 
The linear $M$--$H$ curve at high temperatures becomes \textit{ nonlinear} at 1.6 K as shown in Fig.\,\ref{SQUID}(c) (see also Fig.\,S2 in SM~\cite{SM} for the temperature dependence). 
Below 2~T, $M$ increases faster than the Brillouin curve (the dashed line in Fig.\,\ref{SQUID}(c)). Remarkably, we find that $M$ exhibits a particular field dependence of $M\propto \sqrt{H}$ as shown in the solid lines in Fig.\,\ref{SQUID}(c) and the inset.

\subsection{Magnetic torque measurements}

This $\sqrt{H}$ dependence of $M$ is further confirmed to persist up to 17.5\,T by our magnetic torque measurements done for one single crystal.
Figure~\ref{torque_Hdep}(a) shows the field dependence of the magnetic torque obtained from a fixed-angle high-field torque measurements at 0.12 and 1.7~K. As shown in Fig.\,\ref{torque_Hdep}(a), the magnetic torque shows the field dependence of $H^{3/2}$ (the dashed line in Fig.\,\ref{torque_Hdep}(a)).
Given the form of the magnetic torque $M \times H $, the field dependence of $H^{3/2}$ shows $M \propto \sqrt{H}$.
Note that the free impurity spins are not responsible for this magnetization, 
since otherwise the saturation should take place at $\sim 1$~T for 0.12~K, which is not observed in our data.

\begin{figure}[t!b]
	\centering
	\includegraphics[width=8cm]{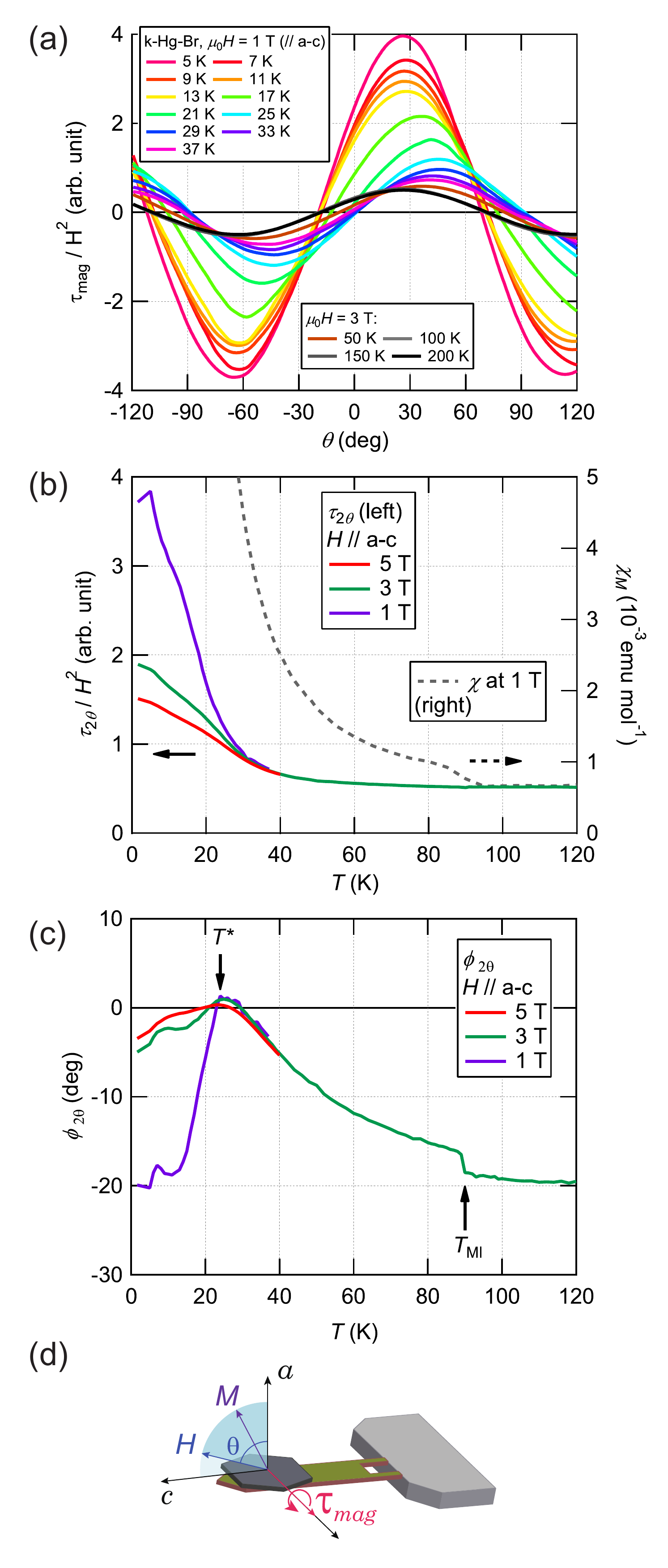}	
	\caption{
		(a) Magnetic torque curves at different temperatures, normalized by $H^2$ to compare the ones for different fields
		(see Section C in SM~\cite{SM} for details). 
		The data below (above) 40~K was measured at 1~T (3~T). 
		(b, c) Temperature dependence of the torque amplitude 
		$\tau_{2\theta}$ divided by $H^2$ (b) and the phase shift $\phi_{2\theta}$ (c) of $\sin 2 \theta$ component of the torque curve. The magnetic susceptibility shown in Fig.\,\ref{SQUID}(a) is also plotted on the right axis as the dashed line in (b). 
		(d) Schematic drawing of magnetic torque measurement by a resistive cantilever. 
	}
	\label{torque_all}
\end{figure}

Throughout the whole sweep of $H$, the $M$--$H$ curve shows neither a remnant field nor a hysteresis (Fig.\,\ref{SQUID}(c)).
The absence of hysteresis is further confirmed down to the lowest fields by our magnetic torque measurements (Fig.\,\ref{torque_Hdep}(b)).
Our data excludes the spin-glass based weak ferromagnetism picture presented in the previous study~\cite{Hemmida2018,fn1}, because both a ferromagnetic state~\cite{Pinteric1999} and a spin glass state~\cite{FertHippert1982} is known to exhibit clear hystereses in the torque measurements from a remnant field and a frozen moment, respectively.

	To investigate the magnetic state below $\TMI$ in detail, we measured the magnetic torque curves by rotating the magnetic field in the $a$-$c$ plane (Fig.~\ref{torque_all}(a)), 
	where $\theta$ denotes the angle between the field and the $a$-axis (see Fig.~\ref{torque_all}(d)). 
	The magnetic torque signal, 
	$\tau_{mag}=\tau_{2 \theta }\sin 2 (\theta - \phi_{2\theta})$, 
	 is obtained after subtracting the $\sin \theta$ component that comes from the gravity of the sample mass 
(see Section C in SM~\cite{SM} for details).
	 These magnetic torque curve measurements allow us to detect the magnitude of the magnetic anisotropy, which is proportional to the amplitude of the $2 \theta$ component divided by $H^2$ (Fig.\,\ref{torque_all}(b)), and the direction of the magnetic principle axis by the phase $\phi_{2\theta}$ (Fig.\,\ref{torque_all}(c)). 
	In the metallic $T > \TMI$ phase, $\phi_{2\theta}$ stays at around $-20$~degree, 
	which is close to the angle between the long axis of BEDT-TTF molecules and the $a$-axis (see Fig.\,\ref{fig1}(a)), 
	showing that the magnetic anisotropy comes from the spins on the BEDT-TTF dimers~\cite{Watanabe2012,Isono2014,Isono2016}.
	
	At $\TMI$, $\phi_{2\theta}$ shows a sharp jump which is followed by a rapid shift of $\phi_{2\theta}$ toward zero, while at the same time $\tau_{2 \theta }$ stays nearly temperature independent in contrast to the increase of $\chi_M(T)$. 
	These contrasting temperature dependences indicate that the magnetic principle axis varies concomitantly with the decrease of the magnetic anisotropy below $\TMI$.
	Since Raman~\cite{Hassan2018} and IR~\cite{Ivek2017} vibration measurements observed no change of the phonon spectrum below $\TMI$, the change of $\phi_{2\theta}$ cannot be attributed to the rotation of the BEDT-TTF molecules. 
	Therefore, this $\phi_{2\theta}$ shift is given by an emergence of a magnetic easy axis parallel to the $a$ axis caused by the ferromagnetic interaction appearing below $\TMI$. 
	A similar but much smaller phase shift has been observed in {\kCN}~\cite{Isono2016}, which may be ascribed to an additional moment from valence bond defects~\cite{Riedl2019}.
	We further find a characteristic temperature $T^* \sim 24$~K.
	Below $T^*$, $\phi_{2 \theta }$ drops, $\tau_{2\theta}/H^2$ increases, and  both $\tau_{2 \theta }/H^2$ and $\phi_{2\theta}$ starts to depend on the field strength. 
	The increase of the magnetic anisotropy particularly developing below $T^*$ is consistent with the 
	anisotropy of $\chi_M(T)$ observed in the previous measurement~\cite{Hemmida2018}, supporting the magnetic origin of these temperature changes. 
	This temperature-dependent change is larger for lower fields; 
	as we saw in Fig.~\ref{torque_Hdep}(b) the torque data at $|H|\lesssim 0.5$~T changes its sign below 20\,K. 
	Another bump-like feature in $\phi_{2\theta}$ is observed around 7\,K, implying a further change of the magnetic state.
	These features might be related to the changes of the relaxation times observed in NMR measurements done at higher fields~\cite{Le2020}.
	We thus observed a distinct change in the magnetic property already starting below $\TMI$ via two torque parameters. 
	Further magnetic torque measurements performed in a dilution refrigerator reveal no change in the magnetic torque below 2~K (Fig.\,\ref{torque_Hdep}(a)), showing a saturation of the temperature dependence.


\section{Discussion}
	
	Our magnetic measurements on {\kHgBr} disclose an unconventional magnetic state, 
	which to our best knowledge has never been observed, 
	in the other family members of $\kappa$-(BEDT-TTF)$_2 X$ -- the ferromagnetic $\TCW$, the non-hysteretic $M$--$H$ curve with $M \propto \sqrt{H}$ at low temperatures, and the large change in the direction of the magnetic principle axis. 

\begin{figure*}[tb]
	\centering
	\includegraphics[width=\linewidth]{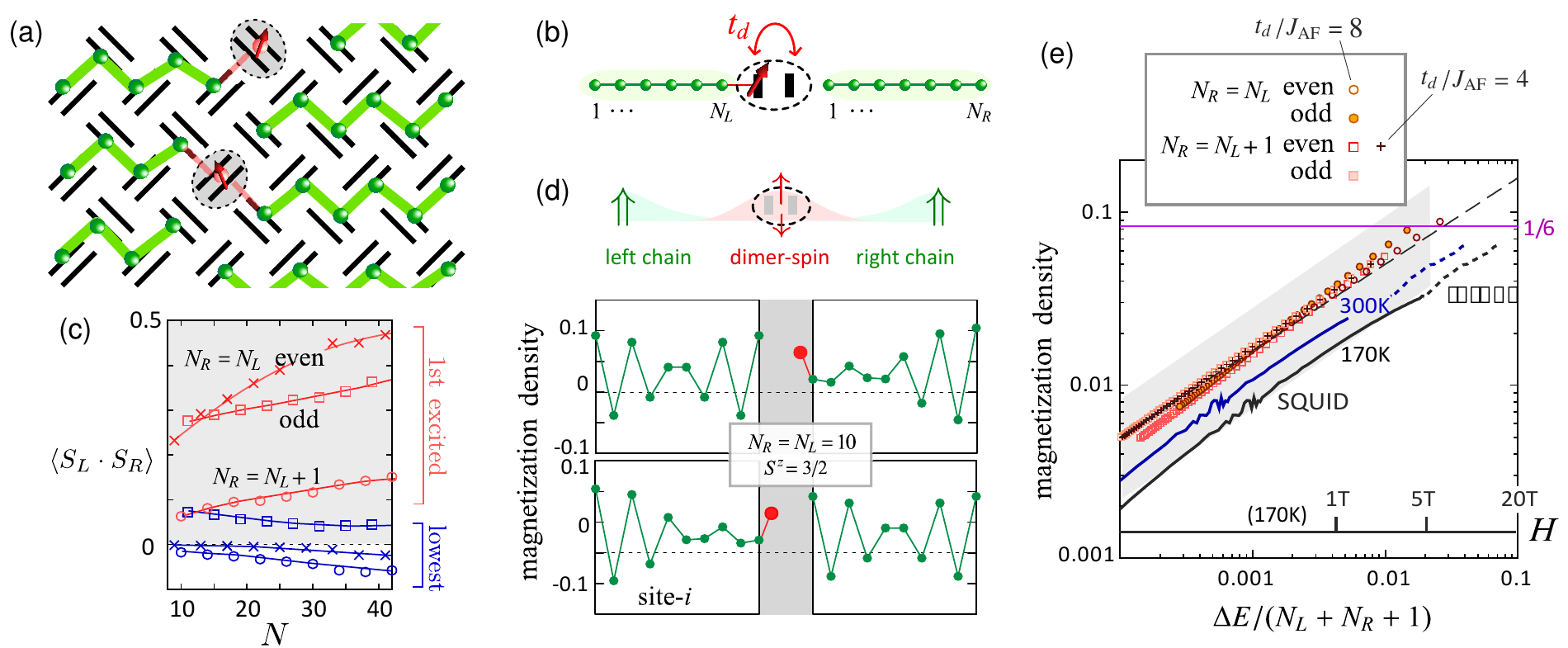}
	\caption{
		(a) One possible charge configuration of {\kHgBr} with a charge-ordered domain separated by the Mott dimer. 
		The ordered charges (green circle) carry spin-1/2 and form an antiferromagnetic spin chain along the green bond. 
		The charge on Mott dimer fluctuates via $t_d$ between dimerized two molecules, and when it occupies one side 
		of the dimerized molecule which finds a green-colored charge on the nearest neighbor site connected 
		by either of $t_B,t_p$ or $t_q$ (see Fig.\,\ref{fig1}(b)), it interacts via $J_{\rm AF}$ colored with red bond. 
		(b) Illustration of the model described by Eq.~(1). Arrow represents the dimer-spin. 
		(c) Correlation $\langle S_L\cdot S_R\rangle$ between spins on left and right chain for several series of $N=N_L+N_R+1$,   			        for the two lowest energy states, 
		with dominant ferromagnetism $\langle S_L \cdot S_R \rangle>0$ (see Section E in SM~\cite{SM}). 
		Red (blue) symbols indicate the data of the 1st-excited (lowest) states. 
		(d) Magnetization density at site-$i$ of the first excited state of Eq.~(1) with $N_L=N_R=10$ and $S^z=3/2$. 
		It consists of the equal weight superposition of upper and lower panels 
		with dimer-spin (in the gray region) on the right and left. 
		(e) Magnetization density of the first excited state of Eq.~(1) as a function of excitation energy $\Delta E$ per spin 
		for $N_L, N_R=8$--100, where the absolute values of the curve may shift within the shaded region when we change 
		the ratio of left and right chains down to $N_R\sim N_L/2$. 
		All give the magnetization curve in a square-root(broken line) form by interpreting the horizontal axis as a magnetic field. 
		The grey and blue bold lines (solid part:SQUID, broken part:torque) are
		the experimental result with vertical axis in unit of $\mu_B/2$, when scaling the horizontal axis as $J_{\rm AF}/k_B \sim 170$\,K and 300\,K as a unit of $\Delta E$, respectively. 
	}
	\label{fig5}
\end{figure*}

	Let us first start by elucidating the way the charges are localized at $T<\TMI$. 
	Most of the previously known $\kappa$-(BEDT-TTF)$_2 X$ become a dimer Mott insulator 
	depicted schematically in Fig.~\ref{fig1}(c). 
	In a Mott phase, the dominant magnetic interactions between the spins carried 
        by the localized charge are always antiferromagnetic 
        as they originate from the kinetic exchange as, 
        $J_{\rm AF}=4t_{ij}^2/(U-V_{ij})$ where $t_{ij}$ and $V_{ij}$ are the transfer integrals 
        and inter-molecular Coulomb interaction along the exchange bond, and $U$ is the on-molecular Coulomb interaction. 
	Then, the antiferromagnetic order of {\kCl} and quantum spin liquid nature of {\kCN} 
	are roughly understood by the square-like and triangular lattice geometry of 
	$J_{\rm AF}$ which amounts to 500~K~\cite{Smith2003} and 250~K~\cite{Shimizu2003}, respectively~\cite{commentJAF}. 
	Therefore, the positive {$\TCW$} observed in {\kHgBr} cannot be explained by the magnetism of a dimer Mott insulator. 
    
        In fact, the abrupt increase of resistivity just below $\TMI$ in both {\kHgBr} and {\kHgCl} is 
        different from the crossover behavior usually observed in dimer Mott materials~\cite{Kanoda2006}, 
        signaling some sort of translational symmetry breaking of charge distribution. 
	However, the Raman spectroscopy measurements indicate the absence of static charge ordering in {\kHgBr}~\cite{Hassan2018}. 
	A scenario compatible with all these findings is the dynamical and inhomogeneous charge distribution in between 
        the dimer Mott and charge-ordered state.       
	The intra-dimer transfer integral from the first-principles calculation on {\kHgBr} is $t_d\!\sim\!120$~meV~\cite{ValentiJeschkePrv,Gati2018}, 
	much weaker than the typical value $\sim 200$~meV 
	of the $\kappa$-salts~\cite{Koretsune2014}, 
	and thus a quasi-charge-order by the inter-dimer Coulomb interactions is a reasonable expectation. 

	In Fig.\,\ref{fig1}(d), we locate {\kHgBr} and {\kHgCl} according to the first-principles based evaluations~\cite{Jacko2020}. 
	In the uniform charge-ordered case possibly realized in {\kHgCl}, 
	$J_{\rm AF}$ forms long quasi-one-dimensional (1D) chains (see the green lines in Fig.~\ref{fig1}(c)). 
        Here the vertical stripe charge configuration is possibly favored for the Coulomb-interaction-strength 
        of $V_q < V_p, V_B$ of the material~\cite{Jacko2020}. 

        When the static and bulk charge order is no longer stabilized in {\kHgBr}, 
        these chains shall break up into short fragments separated by Mott-dimers, as shown in Fig.~\ref{fig5}(a). 
        The way to construct the domain is not really random; we assume that the chain length $N$ roughly corresponds to the 
        correlation length of charges, and a Mott-dimer is inserted between the chains running in the $t_q$ direction, 
        while in reality sometimes there will be a connections with the dimers through $t_p$ in the other directions.
	Inside the 1D fragment the spins interact along the $t_q$-bonds via 
        $J_{\rm AF} \sim 170$--300~K, which will give $t_d/J_{\rm AF}\sim 4$--8 (see Section D in SM~\cite{SM}).
        The charge on a Mott-dimer fluctuates, with fluctuation parameters defined by $t_d$ values. 
        During these fluctuations, the charge (and relevant spin) occupies either left or right molecule on the dimer, 
        and interacts with $S=1/2$(green circle) at the adjacent left/right end of the chain via $J_{\rm AF}$. 
        For the charge configuration shown in Fig.~\ref{fig5}(a), $J_{\rm AF}$ shown in green and red bonds have the same amplitude. 

	To elucidate how these quantum fluctuations modify the dominant antiferromagnetism, 
	we construct a synergetic model~\cite{fn2} consisting of two open chains with 
	$N_{\rm L}$ and $N_{\rm R}$ spin-1/2's and a single electron with $S=1/2$ (which we call dimer-spin) as shown in Fig.~\ref{fig5}(d). 
	The Hamiltonian is given as 
	\begin{align}
	{\cal H}&= \sum_{\gamma={\rm L,R}} 
	\sum_{\langle i,j\rangle} J_{\rm AF} \hat S_{i\gamma} \cdot \hat S_{j\gamma} 
        + t_d \:(c_L^\dagger c_R + c_R^\dagger c_L)\nonumber \\
	&+ J_{\rm AF} \:\big((\hat S_d\cdot \hat S_{N_{\rm L} L}) n_L +(\hat S_d\cdot \hat S_{1R}) n_R\big), 
	\label{1dham} 
	\end{align}
	where $\hat S_{i\gamma}$ is the spin on site-$i$ on left and right chain ($\gamma={\rm L/R}$), 
	$c^\dagger_{\rm L/R}$ and $c_{\rm L/R}$ are the creation and annihilation operator of charges on 
	the left/right molecule of the dimer with its number operator $n_{\rm L/R}=c^\dagger_{\rm L/R}c_{\rm L/R}$, 
        and $\hat S_d$ is the dimer spin. 
	This model cuts out the locally interacting manifold of spins shown in Fig.\,\ref{fig5}(a). 
        Such charge configurations behind the model are expected for {\kHgBr} at temperatures less than $\TMI$,
        where we find no indication of long range order of both charges and spins. 
        The details of electronic state below $T^*$ are not really known, but the present model 
       does not contradict with the experimental reports given so far.

	The model is solved numerically by combining the exact diagonalization calculation~\cite{SM}. 
        Since total-$S^z$ of Eq.\,(\ref{1dham}) is a conserved quantity, we analyze the model by dividing the 
        Hilbert space into different $S^z$-sectors, and evaluating the lowest energy levels for each sector. 
        Figure~\ref{fig5}(c) shows $\langle S_L\cdot S_R\rangle$ between spins on left and right chains, $S_\gamma=\sum_{j\in \gamma} S_j$
        for several different series of $N_\gamma$ and system length $N$. 
        One finds a dominant ferromagnetic correlation ($\langle S_L\cdot S_R\rangle>0$) for large portions of the two lowest excited states. 
        Representative spatial distribution of spin moments for slightly polarized state 
        is shown in Fig.~\ref{fig5}(d); the contribution from the constituents of the wave function 
        with dimer-spin on the right and left are separately drawn. 
	The left-upper panel is a typical spin distribution 
	with two-fold periodic Friedel oscillation generated by the two open edges of the chain~\cite{ShibataHotta2011}. 
        The dimer-spin hops back and forth, 
        and mixes quantum mechanically with spins on closer edges of the chains and suppresses their moments. 
        The moments are redistributed throughout the chains and are accumulated densely on the further edges from the center. 
        They point in the same orientation mediated by fluctuating spins closer to the dimer-spin 
        (top panel of Fig.~\ref{fig5}(d)). 
	This interplay of $t_d$ and $J_{\rm AF}$ generates a robust quantum ferromagnetism (Fig.~\ref{fig5}(e)), 
        which is insensitive to the value of $t_d$ and the choice of $N_\gamma$
        (see Section E in SM~\cite{SM} for details). 
	
	The theory explains the square-root behavior of $M$--$H$ curve at low temperature in {\kHgBr}. 
        Our calculations show that the ground state of Eq.\,(\ref{1dham}) with even $N$ is always 
        nonmagnetic and has dominant antiferromagnetic correlation. 
        Let us consider exciting a magnetic moment by applying a magnetic field.
        Suppose that for an isolated chain with fixed $N_L$ and $N_R$, 
	the lowest eigenenergy of Eq.\,(\ref{1dham}) for each $S_z$ sector is given as $E(S_z)$. 
        In an applied field $H$, the system acquires a finite magnetization $S_z$ that gives the minimum of 
  	energy $E_{H}={\rm min}_{S_z}\big( E(S_z)-S_z H \big)$. 
        The "magnetization curve" at finite $N_L$, $N_R$ is given as such that 
        $H(S_z)= \Delta E/\Delta S_z$, where $\Delta E$ is the energy difference $E(S_z)-E(S_z-1)$ for $\Delta S_z=1$. 
        As mentioned earlier, the magnetism of the short range charge ordered phase shall be described by the assemblage 
        of small magnetic subsystems, 
        interacting with each other, connected with more than two neighboring subsystems.
        Since the information on the distribution of the chain length is missing, 
        and since the calculation is dealing with only two interacting segments,
        the direct comparison of the theory and experiments may seem difficult. 
        However, we find that an unbiased comparison is possible as shown in Fig.~\ref{fig5}(e), 
        where we plotted the magnetization density $S_z/N$ as a function of $\Delta E/N$ 
         of the subsystems with various different $N_L$ and $N_R$. 
	Here, since all the data form a universal square-root curve \textit{regardless of the chain length $N$}, 
        it can be interpreted as a stochastic magnetization curve against magnetic field $H/J_\textrm{AF}$. 
        As found in the logarithmic plot, the functional form, $\Delta E/N \sim C\sqrt{S_z/N}$, always holds 
        regardless of the length of the chains,  
        while the constant $C$ may depend on the ratio of $N_L$ and $N_R$. 
        The universal square-root behavior {\it insensitive to $N$ means that the energy is determined locally}. 
        Accordingly, if we consider a bulk assemblage of segments of chains connected by Mott-dimers, their energy 
        shall be an extensitve quantity, i.e. the summation of local energy gains. 
        Therefore, we consider this functional form to be intrinsic.
      
	The experimental data is plotted together in Fig.\,5(e) where we 
        add the magnetic torque data (data in Fig.\,3(a) divided by $H$) 
        into the field dependence of the magnetization data by SQUID (the data in Fig.\,2(c) as it is) 
        so that $M$ estimated from $\tau_{2\theta}/H$ coincides to $M$ of the SQUID data at 5\,T.
	The horizontal axis of the experimental data is determined by the value of  $J_{\rm AF}$, and is illustrated for the two parameter choices of $J_{\rm AF}/k_B = 170$\,K and 300\,K discussed in Section D of the SM~\cite{SM}.
	The shaded region represents the vertical range over which the absolute value of the magnetic moment may vary
	if the distribution of chain lengths has a large variance, and hence one may state that the theory shows good
	qualitative agreement with the experiment for any comparable choice of $J_{\rm AF}$.

   An extrapolation of the experimental data in Fig. 5(e) shows that the magnetization reaches $\mu_B/6$ at about 20--30\,T.
        Therefore, approximately, the field strength of 20--30\,T which is comparable to $T^*=24$\,K, 
        gives the energy scale to excite the 1/6 magnetic moment from the nonmagnetic ground state. 
        At the temperature range $T^* < T< \TMI$, such 1/6 moment is thermally excited and contributes to the ferromagnetism; 
        the ferromagnetically coherent orientation of the moment would contribute to the phase shift of $\phi_{2\theta}$. 
        There, $M$--$H$ curve no longer has a square-root, because the low energy magnetic excitations are smeared out.
        The energy scale of $\mu_B H \sim 0.1 J_{\rm AF}$ to have the $\mu_B/6$ moment is consistent with 20--30\,T.

	Also, the preserved SU(2) symmetry in Eq.\,(1) matches with 
        the restored isotropy in the magnetic torque at $T<\TMI$. 
	Notice that this ferromagnetic phase is not a long-range order but a correlation because of the one-dimensionality, 
	as can also be suggested from the lack of the hysteresis.         
	Below $T^*$ the nonmagnetic ground state component becomes dominant. 
	From Raman spectroscopy measurements, the static charge ordering is excluded, 
        whereas the broad peak in $\nu_2$ mode is still compatible with quasi-charge-ordered domains 
        with a variant charge disproportionation maximally amounting to $\pm 0.1e$, 
        which are coherently fluctuating together inside the domain with a frequency estimated as 1.3~THz~\cite{Hassan2018,Ivek2017}.

	While evaluating the precise character of the charge distribution is beyond the scope of any theory currently available, in Section F of SM~\cite{SM} we provide a phenomenological treatment performed by assuming a functional form for $\xi(T)$ that is valid throughout the critical regime. Within this approach, we show that $\chi$ manifests a	Curie-Weiss-like behavior that reflects the ferromagnetic correlations between thermally excited spins at temperatures $T>T^*$. Further experimental information concerning the functional form of $\xi(T)$ is required to verify this type of treatment.

        Once the temperature falls below $T^*$, the dipole (charge) degrees of freedom become correlated over a length scale $\xi$ whose $T$-dependence saturates, and fluctuates slowly together at a corresponding timescale. These fluctuations can safely be integrated out (see Ref.\,\cite{Hotta2010}), leading to the effective model of Eq.\,(1) for spins with antiferromagnetic interactions on chain segments of average length $\xi$.

	\par
	One may suspect that the spin models with extrinsic impurities can also explain the phenomena. 
        Although the possibility of spin glass is experimentally excluded, 
        its quantum analogue, the random singlet phase may share similar feature with the present magnetism~\cite{Shimokawa2015}; 
        most of the spin moments form a singlet and the remaining spins may contribute to the magnetism. 
        However, for such state to happen one needs a large amount of static randomness in the distribution of 
        $J_{\rm AF}=J(1\pm \Delta)$ that amounts to $\Delta\gtrsim 0.6$~\cite{Shimokawa2015}, 
        which cannot happen in the present system. 

	\par
	Ferromagnetism is elusive; 
	For molecular-based materials with only light elements, 
	few ferromagnetic compounds are known, e.g. p-NPNN~\cite{Takahashi1991}, 
	C$_{60}$(TDAE)$_{0.86}$~\cite{ALLEMAND1991}, and (Et-4RrT)-[Ni(dmit)$_2$]$_2$~\cite{Kusamoto2013}, 
	whose exchange interactions originate from the higher-order Goodenough-Kanamori rule. 
	Other mechanisms of bulk ferromagnetism known so far are the 
	Nagaoka ferromagnetism~\cite{Nagaoka1966}, flat-band~\cite{Tasaki1998}, 
        and double exchange or multi-orbitals Hubbard models~\cite{Zener1951,Kubo1972,Sakamoto2002}, which are applied to metals. 
	The present finding should thus be the first proposal of generating a robust ferromagnetic exchange 
	from the inhomogeneous charge distribution forming dominant spin singlet (paramagnet) formed by 
        the leading antiferromagnetism interactions. 

	Naively, our ferromagnetism can be viewed as a local double-exchange; 
        a single charge hops back and forth inside the dimer, and since it interacts with the spin chains 
        on both sides, it is favorable to have both point in the same direction to maximally gain 
	the antiferromagnetic exchange interaction. 
        The difference from the double-exchange is that the kinetics of charge is local, keeping the insulating character, 
        and the weakly coupled one-dimensional fragments of spin chains do not allow the development 
        of coherent ferromagnetic long range order. 
	
	By designing a three-dimensional critical phase in between the charge order and dimer Mott insulator 
        a two dimensional ferromagnet might be available, 
        in which case the ferromagnetic long-range order is allowed at a finite temperature. 

\bigskip
	\noindent
	{\bf Materials and Methods:}
	Single crystals of {\kHgBrFull} were grown by electrochemical oxidation of the BEDT-TTF solution~\cite{Hassan2018}.
	In this synthesis, many single crystals with a variety of the size were obtained. We used a batch of the smaller crystals 
	(powder samples, 3.73\,mg) for the SQUID measurements and picked up larger ones for the magnetic torque measurements.
	
	The magnetic susceptibility of powder samples was measured by SQUID for 1.6--300\,K. 
	The magnetic torque measurements were carried out for one single crystal with dimensions 0.75$\times$0.57$\times$0.13\,mm$^3$ attached to a piezo-resistive cantilever by a tiny amount of grease (see Fig.\,S3(a) in SM~\cite{SM}) by using a variable temperature insert (1.6--200\,K) and a dilution refrigerator (0.09--2.5\,K). 
	The magnetic torque ($\bm{\tau}_{mag} = \bm{M} \times \bm{H}$) is measured by the change of the resistance of the piezo-resistive cantilever.
	The temperature dependence of the sensitivity of the piezo-resistive cantilever is calibrated by the $\sin \theta$ component in the torque curve by the gravity of the sample mass as described in the section C in SM~\cite{SM}. The standard deviations of all the data shown in the figures are smaller than the symbol size. 
	
\bigskip
\noindent
{\bf Data and Code Availability:}
All the data and the numerical codes that support the findings of this study are available from the corresponding author upon reasonable request.
	
\bigskip
\noindent
{\bf Acknowledgements:}
		The authors thank M. Urai  for fruitful discussions.
		The work in JHU was supported as part of the Institute for Quantum Matter, an Energy Frontier Research Center funded by the U.S. Department of Energy, Office of Science, Office of Basic Energy Sciences under Award Number DE-SC0019331.
		The work in Chernogolovka was carried out within the state assignment 
		(number AAAA-A19-119092390079-8).
		The work in Japan was supported by KAKENHI (Grants-in-Aid for Scientific Research)
		Grants No. JP17K05533, No. JP18H01173, No. JP17K05497, No. JP17H02916, No. JP18H05225, No. JP18H05516, No. JP19K05397, No. JP19H01848, and No. JP19K21842.

\bigskip
\noindent
{\bf Author contributions:}  M.Y., H.M. and N.D. conceived the project.
E.Z., S.T., and R.L. prepared the single crystals.
A.U., S.D., Y.S., and H.M. performed the magnetic susceptibility measurements.
M.Y., S.S., T.T., S.U, and N.D. performed the magnetic torque measurements.
C.H. is responsible for the whole theoretical part. 
M.Y., N.D., and C.H. wrote the manuscript.
All the authors discussed the results.

\bigskip
\noindent
{\bf Competing interests:}  The authors declare that there are no competing interests.

\newpage

\renewcommand{\theequation}{S\arabic{equation}}
\setcounter{equation}{0}
\renewcommand{\thetable}{S\arabic{table}}
\setcounter{figure}{0}
\renewcommand{\thefigure}{S\arabic{figure}}

\begin{titlepage}
	\begin{center}
		\vspace*{12pt}
		{\Large Supplemental Material for ``Ferromagnetism out of charge fluctuation of strongly correlated electrons in \kHgBrFull"}
		\vspace{12pt} \\
		\begin{tabular}{c}
			Minoru Yamashita, Shiori Sugiura, Akira Ueda, Shun Dekura, Taichi Terashima,\\
			Shinya Uji, Yoshiya Sunairi, Hatsumi Mori, Elena I. Zhilyaeva, \\
			Svetlana A. Torunova, Rimma N. Lyubovskaya, Natalia Drichko, and Chisa Hotta
		\end{tabular} \vspace{3pt} \\
	\end{center}
\end{titlepage}

\subsection{Comparison of $\chi_M (T)$ data with the previous data}

\begin{figure}[t!b!h!p]
	\includegraphics[width=\linewidth]{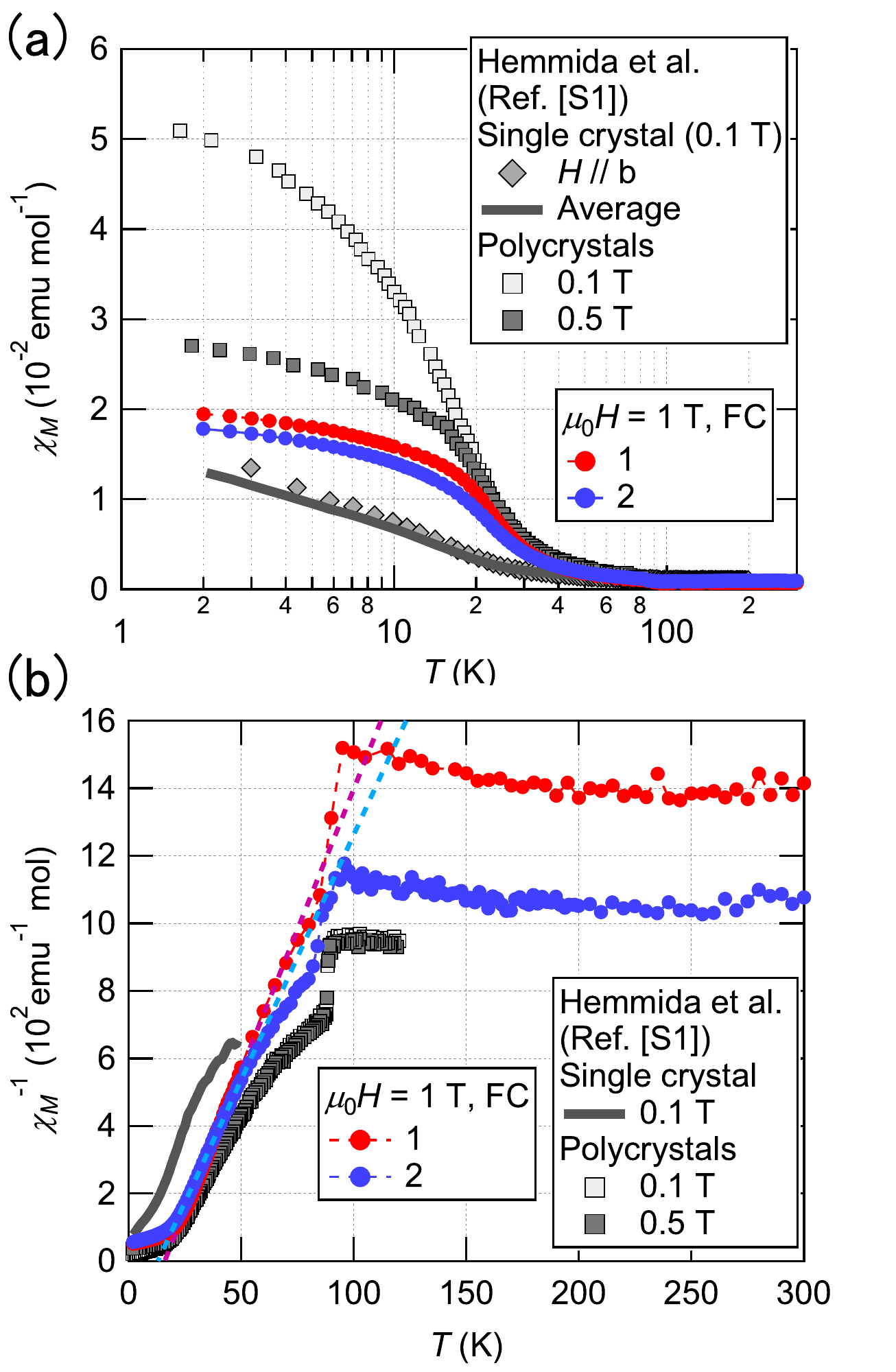}
	\caption{The temperature dependence of the magnetic susceptibility (a) and the invserse of the magnetic susceptibility (b) of our data (filled circles) shown with the data in the previous report taken from Fig.~4 (single crystal, grey diamonds) and Fig.~5 (polycrystals, grey squares) in Ref.~\cite{Hemmida2018}. 
		For the single crystal data from Fig.~4 in Ref.~\cite{Hemmida2018}, we averaged the data taken for $H \parallel a$, $b$, and $c$ to compare the data of polycrystals.
		Only the data obtained by field cool (FC) are shown.
	}
	\label{SQUID_SM}
\end{figure}

Figure~\ref{SQUID_SM} shows the data of the magnetic susceptibility $\chi_M (T)$ of Fig.~2 in the main text (red circles), that of sample 2 (blue circles), and those from the previous report~\cite{Hemmida2018} consisting of two different series of data (grey line and squares).
To compare to our polycrystal results of $\chi_M(T)$, the single crystal data of previous report~\cite{Hemmida2018} is averaged for all axes, showing a large sample variance in their results.
A Curie-Weiss fit of sample 2 (the blue dashed line in Fig.~\ref{SQUID_SM}(b)) also gives a positive Curie-Weiss temperature $\TCW \sim 13$~K with the Curie constant $C=0.069$~emu~K~mol$^{-1}$, showing a good reproducibility of our result of sample 1.
As shown in Fig.~\ref{SQUID_SM}(b) (the dashed lines), our results for both samples show a positive Curie-Weiss temperature more clearly in a Curie-Weiss fitting for a wider temperature range.
A positive Curie-Weiss temperature can also be deduced from the polycrystal data of Ref.~\cite{Hemmida2018} by fitting in a narrower temperature range at lower temperatures.
Instead, Ref.~\cite{Hemmida2018} focused on the field dependence of their $\chi_M (T)$ data only 
in the limited range below $\sim 20$~K off the Curie-Weiss region, 
and together with the results from the ESR measurements, 
argued that there exists a spin glass state with a weak ferromagnetic moment. 
In our case, we confirmed a reproducibility of our torque measurements done in other samples for different magnetic field orientations, confirming that there is no intrinsic sample dependence in our results. 

\subsection{Temperature dependence of the $M$--$H$ curve at high temperatures}

\begin{figure}[t!b!h!p]
	\includegraphics[width=\linewidth]{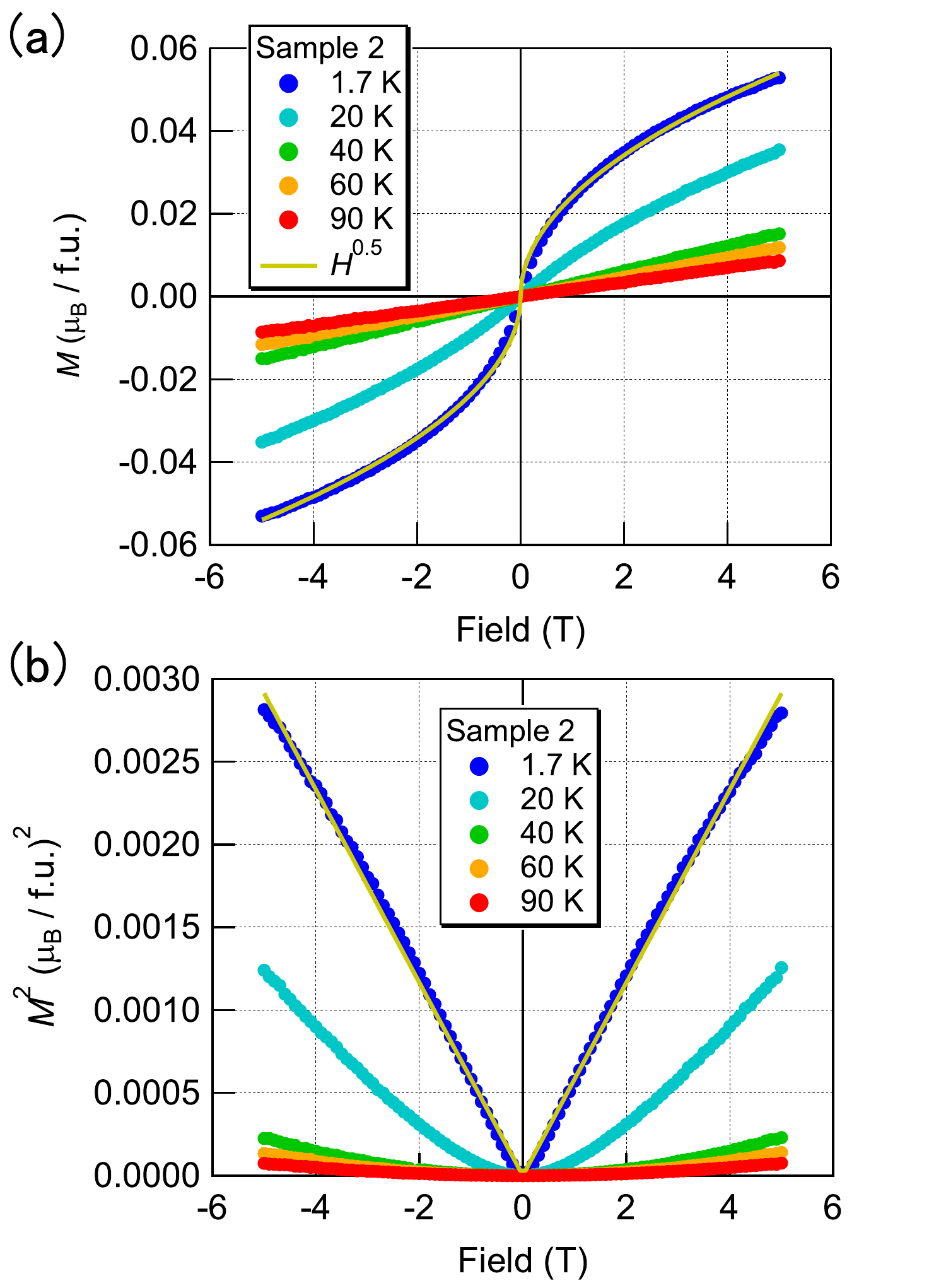}
	\caption{(a) The field dependence of the magnetization ($M$) of sample 2 at 1.7, 20, 40, 60, and 90\,K. 
		The solid line shows a fit to $M \propto \sqrt{H}$.  
		(b) The field dependence of $M^2$ of the same data shown in (a).
	}
	\label{SQUID_MH}
\end{figure}

The temperature dependence of the $M$--$H$ curve of sample 2 is shown in Fig.\,\ref{SQUID_MH}(a).
As shown in Fig.\,\ref{SQUID_MH}(a), the $M$--$H$ curve is linear above 20\,K, which becomes non-linear at lower temperatures.
At 1.7\,K, the field dependence of $M$ is well fitted to $M \propto \sqrt{H}$ (see the solid lines in Fig.\,\ref{SQUID_MH}(a) and (b)), showing a good reproducibility of the sample 1 data shown in the main text.

\subsection{Calibration of the piezo-resistive cantilever by the gravity signal}\label{sec:calib}

\begin{figure}[t!b!h!p]
	\includegraphics[width=\linewidth]{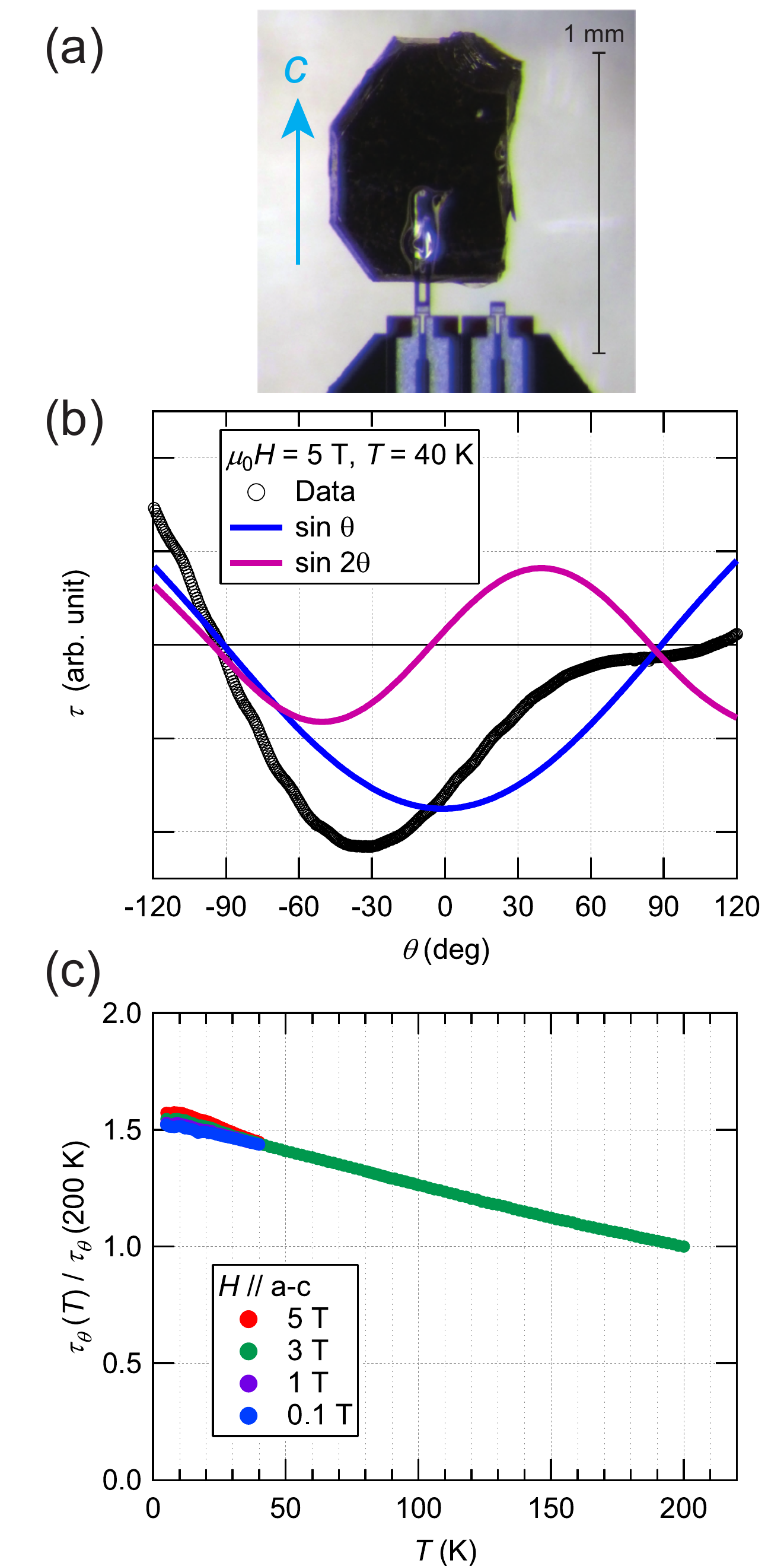}
	\caption{(a) A picture of the single crystal with dimensions 0.75$\times$0.57$\times$0.13\,mm$^3$ attached to the piezo-resistive cantilever for the magnetic torque measurements. 
		(b) A torque curve measured at 5~T and 40~K. 
		The angle $\theta$ is determined by the angle between the magnetic field and the $a$ axis (see Fig.~4(d) in the main text).
		The data (black circles) is given by a sum of the $\sin \theta $ (blue line) and the $\sin 2\theta $ (pink line) components.
		(c) The temperature dependence of the amplitude of the $\sin \theta$ component.
		The data is normalized by that at 200 K.
	}
	\label{torque_gravity}
\end{figure}

In this section, we explain how we calibrated the temperature dependence of the sensitivity of the torque cantilever by using the gravity signal.
We measured the angle dependence of the torque by rotating the sample attached to a piezo-resisitive cantilever (Fig.~\ref{torque_gravity}(a)) in a magnetic field.
The torque signal is given by
\begin{equation}\label{eq_trq}
	\tau (\theta) = \tau_{\theta} \sin (\theta - \phi_{\theta}) + \tau_{2 \theta} \sin 2(\theta - \phi_{2\theta})\,.
\end{equation}
The first $\sin \theta$ term represents the gravity torque coming from the sample mass, and the second $\sin 2\theta$ term represents the magnetic torque ($\bm{\tau}_{mag} = \bm{M} \times \bm{H}$).
Figure~\ref{torque_gravity}(b) shows a typical torque curve which consists of the gravity torque (the blue line) and the magnetic torque (the pink line).
As shown in Fig.\,\ref{torque_gravity}(b), the different oscillation frequency allows one to clearly separate these two signals.
The accuracy in the estimation of the magnetic torque signal is limited by the noise of the torque signal itself ($\sim3$\%), which is mainly caused by the irregular motions of the rotator.

The temperature dependence of the gravity signal (Fig.~\ref{torque_gravity}(c)) reflects the temperature dependence of the sensitivity of the piezo-resistive cantilever, which is used to calibrate the magnetic torque signal obtained at different temperatures.
We note that, although the gravity signal shows a small field dependence at low temperatures (up to $\sim 3$\% of the data), the ambiguity owing to this field dependence is so small (comparable to the symbol size of the plot) that the field dependence can be safely ignored.

\subsection{Analysis on the effective model Eq.\,(1)}
We analyze the effective Hamiltonian Eq.\,(1) in the main text which we rewrite here: 
\begin{align}
	{\cal H}&= \sum_{\gamma={L,R}} 
	\sum_{\langle i,j\rangle} J_{\rm AF} S_{i\gamma} \cdot S_{j\gamma} + t_d (c_L^\dagger c_R + c_R^\dagger c_L)\nonumber \\
	&+ J_{\rm AF} \big((S_d\cdot S_{N_{\rm L} L}) n_L +(S_d\cdot S_{1R}) n_R\big). 
	\label{1dham} 
\end{align}
We consider two quantum spin chains consisting of $N_L$ and $N_R$ sites, 
where the spins on one edge of both chains 
can interact also with the adjacent dimer-spin $S=1/2$ 
when it is on the left and right side of the dimerized two molecules. 
The antiferromagnetic interaction, $J_{\rm AF}$ is evaluated as 
$J_{\rm AF} \sim 4t_q^2/(U-V_q)$ where $t_q$ is the transfer integral connecting the green bond 
with index-$q$ in Fig.~1(b), 
and $U$ and $V_q$ are the on-molecule and inter-dimer Coulomb interaction, respectively. 
\par
The model parameters of the materials are evaluated based on the first principles calculation. 
First, we consider as a reference a dimer Mott insulator, {\kClFull} and {\kCN}, 
where spin-1/2 is localized on each dimer, forming a quantum spin-1/2 Heisenberg system on a triangular lattice.
The fit of the experimentally measured susceptibility by the high-temperature expansion 
gives $J_{\rm AF} \sim 500$~K~\cite{Smith2003} and $\sim 250$~K~\cite{Shimizu2003}, respectively. 
Independently, from the inter-dimer transfer integral $t$, 
one can evaluate $J_{\rm AF}= 4t^2/U_{\rm dimer}$ with $t\sim 70$~meV for {\kClFull} 
and 50~meV for {\kCN}~\cite{Koretsune2014}, 
which gives the above experimentally derived values if we take $U_{\rm dimer}\sim 460$~meV. 
Simultaneously, from the first principles and {\it ab initio} and cRPA study~\cite{Nakamura2009} 
giving $U\sim 0.83$\,eV $V\sim 0.4$\,eV, and $t_d\sim 200$~meV, 
and using the formula (see Ref. [12] in the main text), 
\begin{equation}
	U_{\rm dimer}= 2t_d + \frac{U+V_d}{2}-\sqrt{\frac{(U-V_d)^2}{4} + 4t_d^2}, 
	\label{udimer}
\end{equation}
we find $U_{\rm dimer}\sim 0.65$\,eV. 
Therefore, in reality the Coulomb interaction effect has ambiguity and may be properly reduced by about 
30--40\%
from the {\it ab initio} and cRPA values. 
Here, notice that previously the Coulomb interaction on an isolated dimer was evaluated as $U_\textrm{dimer} \sim 2 t_d$, 
and since $t_d$ differs much between materials, so is $U_\textrm{dimer}$ in that context. 
However, this evaluation is valid in the limit of very large $U$ and $V=0$ in Eq.\,(\ref{udimer}), 
which is an unrealistic situation. 
Recent theoretical studies~\cite{Koretsune2014,Nakamura2009} revealed that $U_\textrm{dimer}$ does not depend much on materials, because the face-to-face distances between dimerized molecules, and $U_\textrm{dimer}$ 
is insensitive to the relative angles between molecules (unlike $t$).

\begin{figure}[tbp]
	\includegraphics[width=8cm]{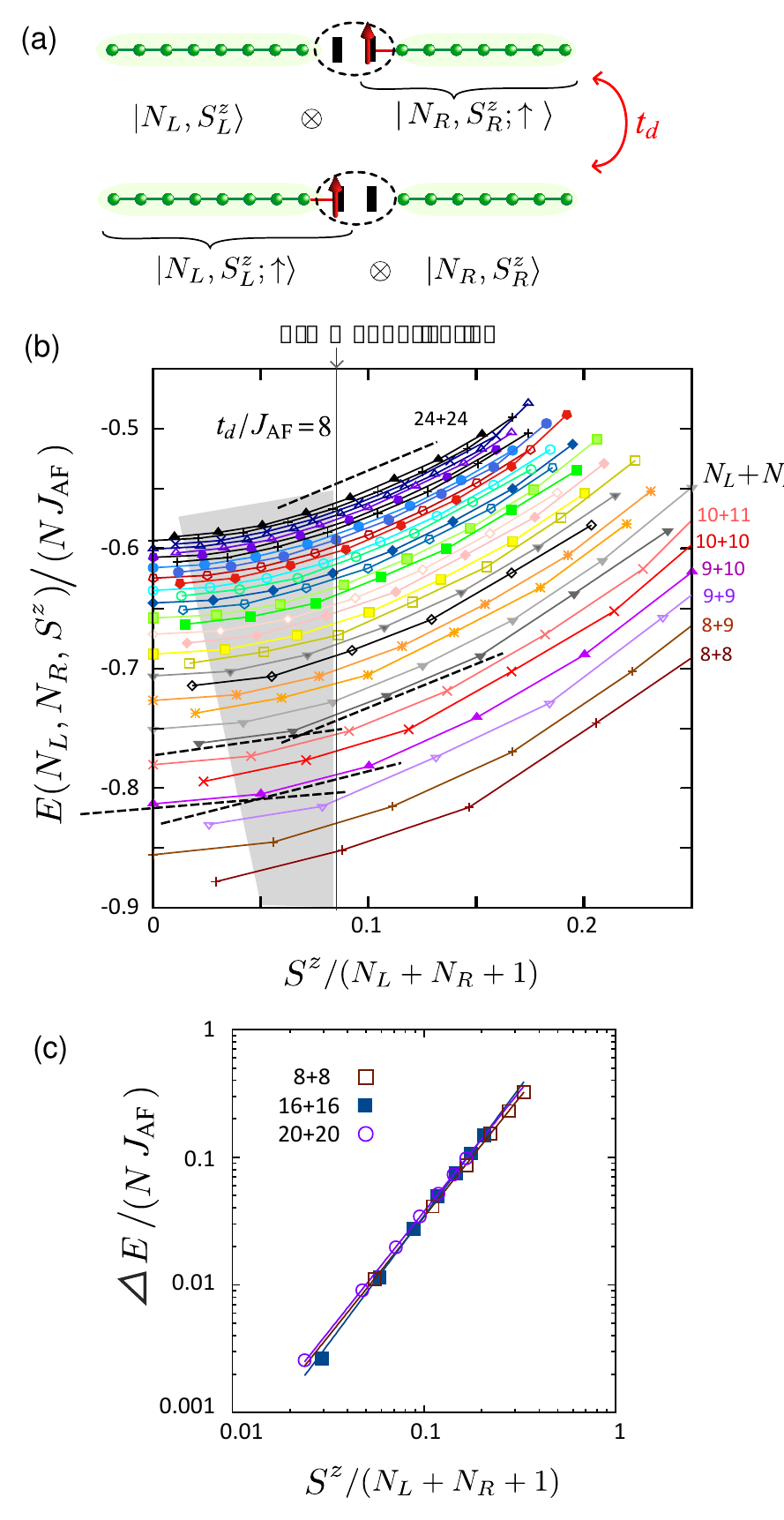}
	\caption{(a) Schematic illustration of two configurations. 
		The upper and the lower panel shows a up dimer spin belonging to the right and left chain, respectively. 
		They are part of the low energy basis of Eq.\,(\ref{1dham}). 
		(b) Lowest eigenenergy $E(N_L,N_R)$ of Eq.\,(\ref{1dham}) for given sets of $N_L=N_R=6-20$. 
		For odd and even $N=N_L+N_R+1$, 
		the lowest magnetization (starting point of the data) is 1/2 and 0, respectively. 
		The energy levels in the shaded region are {\it the magnetically excited states} with 
		$S^z$-density being less than 1/6 of the full magnetization, 
		which are approximately the ones that contribute to the ferromagnetism. 
		Broken lines are the gradient of the data that roughly gives the magnetic field strength 
		required to magnetize the chain up to that point. 
		(c) $\Delta E=E(N_L,N_R,S^z)-E(N_L,N_R,S^z=0 \:{\rm or}\:1/2)$ for $(N_L,N_R)=$(8,8), (16,16), and (20,20), 
		divided by $NJ_{AF}$, using the same data from panel (b).
		\vspace{10mm}
	}
	\label{fS1}
\end{figure}
\par
If we also adopt $U\sim 0.83$\,eV and $V_q\sim 0.4$\,eV in our material {\kHgBr}, 
and use the first principles results $t_q=40$~meV \cite{John2018}, 
we find $J_{\rm AF} = 4t_q^2/(U-V_q) \sim 15$~meV $\sim$ 170~K. 
Here, $U-V_q$ is the energy difference between 
the Mott state and the excited state that has doubly electron-occupied molecule, 
where we set $(U - V_q)\sim 430$~meV.
If we reduce the Coulomb interaction energy in the numerator by 40$\%$, the value will become $J_{\rm AF}\sim$ 290~K. 
Notice that our evaluation does not agree with $\sim $70~K in Ref.\,[\onlinecite{Hemmida2018}]
estimated using magnetization data between 90 and 50 K, and fitting them with negative Curie $T$, 
which is not enough precise because it depends on the fitting range. 
The inter-dimer transfer integral is $t_d=126$~meV from the same first-principles evaluation. 
Based on this consideration, we take $J_{\rm AF}=170$--300\,K, $t_d/J_{\rm AF}\sim 4$--8 which we adopt in the following. 
Our numerical results remain almost quantitatively unchanged by the variation of $t_d/J_{\rm AF}$ within this range. 

\par
The model (\ref{1dham}) is solved in a two-fold manner. 
We first diagonalize the Hamiltonian of a simple spin chain of length $N_\gamma$ with open boundary given as 
\begin{equation}
	{\cal H}_{\gamma}=\sum_{j=1}^{N_\gamma-1} J_{\rm AF} S_{j} \cdot S_{j+1}, 
\end{equation}
in unit of $J_{\rm AF}=1$ and obtain few lowest eigeneneriges $\epsilon_n(N_\gamma, S^z_\gamma)$ ($n=1,2,3\cdots$) 
and $|N_\gamma, S^z_\gamma\rangle_n$, 
for each fixed value of the quantized $z$-component of total spin, $S^z_\gamma$ 
(which is integer/half-integer for even/odd $N_\gamma$). 
Along with this, we elso prepare a set of eigenstates, 
$\big\{ |N_\gamma, S^z_\gamma; \uparrow \rangle_n \big\}$ and $\big\{ |N_\gamma, S^z_\gamma; \downarrow \rangle_n \big\}$, 
of ${\cal H}_{\gamma}^\sigma={\cal H}_{\gamma}+ J_{\rm AF} S_{N_\gamma \gamma}^z \cdot S^z_{d}$, 
where the dimer-spin $S^z_d=\uparrow, \downarrow$ is attached to one edge of the spin chain. 
By using these low energy eigenstates as building blocks 
one can construct the low energy basis of Eq.\,(\ref{1dham}). 

%
In Eq.\,(\ref{1dham}), the total $S^z=S^z_L+S^z_d+S^z_R$ of the whole system of size $N=N_L+N_R+1$ is a conserved quantity, 
so that its low energy basis is a combination of different choices of 
$(S^z_L, S^z_d, S^z_R)$ in each total $S^z$-sector. 
Also, depending on whether the dimer-spin $S_d$ is interacting with the left or right chain, 
the basis includes a variety of states. 
The off diagonal terms of Eq.\,(\ref{1dham}) between these basis are given for example as, 
\begin{eqnarray}
	&& |N_L,S^z_L;\uparrow\rangle \otimes |N_R, S^z_R\rangle t_d 
	\langle  N_L,S^z_L |\otimes \langle N_R, S^z_R ;\uparrow|, 
	\nonumber\\ 
	&& |N_L, S^z_L;\uparrow \rangle \otimes |N_R, S^z_R\rangle \frac{J_{\rm AF}}{2} 
	\langle N_L,S^z_L+1;\downarrow |\otimes \langle N_R, S^z_R |, 
	\nonumber\\
\end{eqnarray}
which are partly displayed in Fig.~\ref{fS1}(a). 
By diagonalizing the representation of Eq.\,(\ref{1dham}) spanned by the low energy basis we obtain the eigenstates 
as superpositions of these basis states. 
We denote the energy and eigenstate as $E_n(N_L, N_R,S^z)$ and $|N_L,N_R,S^z\rangle_n$, 
$n=0,1,2\cdots$ for each given $S^z$ sector.

\begin{figure*}[t!b!]
	\includegraphics[width=17cm]{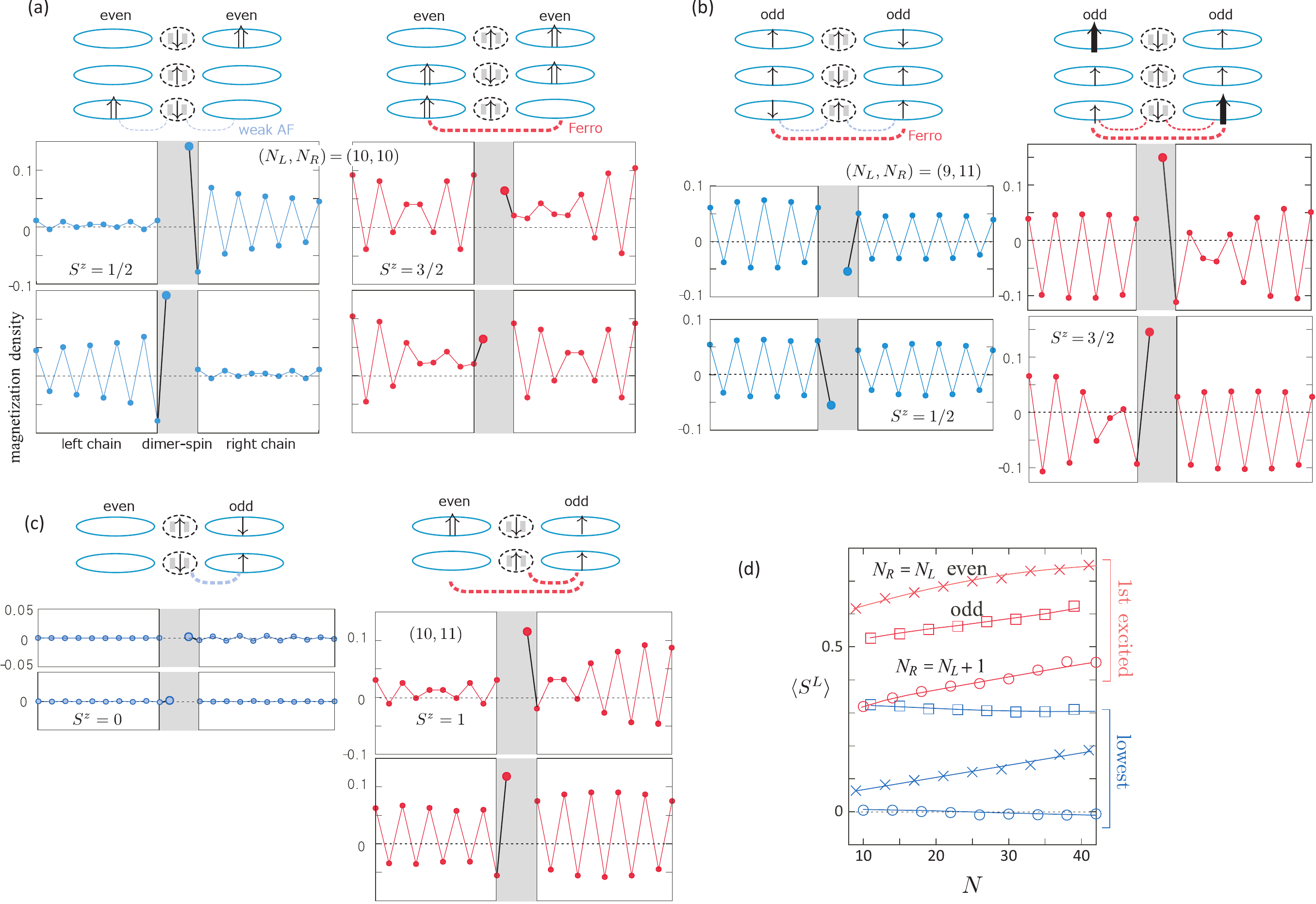}
	\caption{(a--c) 
		Schematic illustration of the spin moments on three different parts of the system, 
		left chain, dimer-spin, and right chain, written in ovals, 
		in the major basis states that contribute to 
		the two lowest energy states of the model in Eq.\,(\ref{1dham}). 
		The thin, double and bold arrows represent the spin-1/2, 1, and 3/2, respectively. 
		Blue and red colors of the symbols classify the lowest and second-lowest energy states, respectively. 
		Dimer spins fluctuate left and right for each depicted spin configurations, 
		and the linear combination of these sketches form the quantum mechanical state. 
		Broken lines are the types of correlations that develop between the spins on these three parts. 
		The lower two panels are the magnetization density obtained by the actual calculations 
		for given $N_L, N_R$ and $S^z$, 
		where the values are separately calculated in the upper and lower panels 
		for the two groups of basis with dimer-spin on the right and left, respectively. 
		(d) Expectation values of magnetization $\langle S_L\rangle$ of spins on a left chain for different $N=N_L+N_R+1$ 
		with $N_L=N_R$ and $N_L+1=N_R$ for the two lowest energy states. 
		The corresponding spin correlations $\langle S_L\cdot S_R\rangle$ are displayed in Fig.~5(c). 
	} 
	\label{fS2}
\end{figure*}

\subsection{Magnetic properties of the effective model Eq.\,(1)}

\begin{figure*}[b!tp]
	\includegraphics[width=15cm]{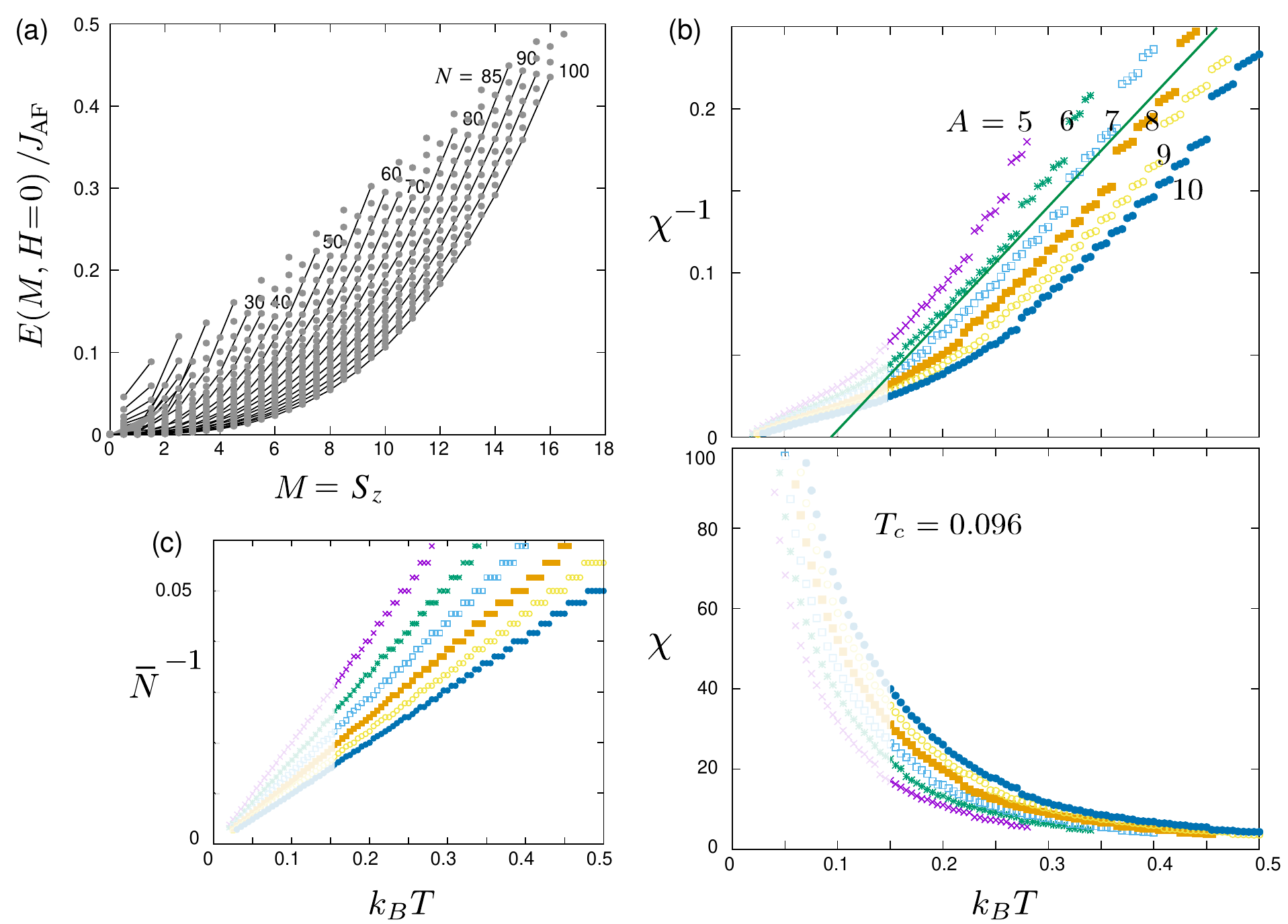}
	\caption{
		(a) Energy for the length-$N$ chain with magnetization $M$ at $H=0$, 
		derived from the fitting to the curve Fig.\,5(e) 
		and the phenomenological calculation based on the square-root $M$-$H$ curve. 
		We plot the energy levels for $N=9$ to 100 for $M=0$(even-$N$) or 1(odd-$N$) up to $N/6$.
		The data obtained in the same $N$ are connected by the solid line. 
		(b) Susceptibility $\chi$ obtained by assuming $A=5$--10 and $\sigma=10$ in Eq.\,(\ref{prob}). 
		Solid line is the guide to the eye that gives $\chi^{-1}\propto (T-T_c)$ with $T_c=0.096$ 
		in unit of $J_{\rm AF}=1$, 
		which corresponds to the experimental evaluation, $T_c=16$\,K, from the Curie-Weiss fit. 
		(c) The mean chain length $\bar N$ for a given distribution function $P(N)$, which we adopted 
		in obtaining $\chi$ in panel (b), following $\bar N=A/k_B T$.
		In the three plots in panels (b) and (c), the region $k_BT\lesssim 0.15 J_{\rm AF}$ are shaded since 
			the assumption we made in obtaining  these figures become no longer valid. 
			Namely, we may no longer expect the growth of the length of chains at this temperature range, 
			which should be of the order of charge-charge correlation length, $\xi$.   
			As we discussed in the main text, $\xi$ should saturate in reality in the actual material 
			at $T < T^* \sim 0.15 J_{\textrm AF}/k_B$.
	}
	\label{fS3}
\end{figure*}

When the two spin chains are disconnected from the dimer-spin, 
the lowest energy state of even-$N_{L/R}$ chain is a singlet, and the first excited state carries spin-1. 
For odd-$N_{L/R}$ chain, the lowest energy state already hosts spin-1/2. 
When they are coupled by the dimer-spins in Eq.\,(\ref{1dham}), 
the lowest energy state based on these singlets still remains almost nonmagnetic, 
whereas all the states based on states with finite magnetic moments on both chains have robust ferromagnetic correlations, 
which we explain here in more detail. 

Figure~\ref{fS1}(b) shows $E(N_L, N_R, S^z)$ for several chices of $N_L=N_R=8$--24 chains at $t_d/J_{\rm AF}=8$. 
The lowest $S^z$ starts from 0 and 1/2 for even and odd $N=N_L+N_R+1$, respectively. 
As mentioned above, the smallest $S^z=0$ or 1/2 has the lowest energy, 
$E(S^z=0\,{\rm or}\,1/2)$ or for each chain length. 
While the shorter chain has lower $E(S^z=0\,{\rm or}\,1/2)$, it does not mean that 
the shorter $N$'s are realized, because the length of the chain is determined in advance by the energetics of the charge degrees of freedom. 
Here, we are interested in the magnetic excitation energy 
$\Delta E=E(N_L, N_R, S^z)-E(N_L, N_R, S^z=0\, {\rm or}\,1/2)$. 
We mark as shaded region above the lowest energy level up to the 1/6 concentration of the full moment. 
As shown for the selected three sets of $(N_L, N_R)$ in Fig.~\ref{fS1}(c), 
$\Delta E/N$ follows a universal functional form. 
Since the derivatives of $\Delta E$ against the excited magnetic moment $\Delta S^z$ gives the 
$M$--$H$ curve, we also find a universal curve in Fig.~4(e), which follows a square-root behavior. 

\par
Let us examine these low energy states by classifying them to three different groups; 
($N_L$, $N_R$) consisting of the combination of (even, even), (odd, odd) and (even, odd) numbers, 
and for each of them we examine the lowest and second-lowest energy states. 

Figure~\ref{fS2}(a)--(c) show the spatial distribution of magnetization density, 
where the upper and lower panels are the ones separately calculated for the groups of basis that have dimer-spin on the right and left part of the dimer, respectively. 
(Fig.~\ref{fS1}(a) left panel is the same as Fig.~5 in the main text). 
The wave function consists of the anti-bonding superposition of these two groups of the basis of equal weight. 
By further examining the composition of the basis one can simply depict the major configurations that have dominant contributions, 
which we show schematically on the upper part of these panels. 
For example, in (even,even) chain the lowest energy state $S^z=1/2$ (left part of (a))
consists of linear combination of three manifolds of states, 
$|0,\downarrow,\Uparrow\rangle$, $|0,\uparrow,0\rangle$, $|\Uparrow,\downarrow,0\rangle$, 
where the three arrows/0 indicate the spins that are carried by the three parts of the system; 
left, dimer, and right chain, 
and thin $\uparrow/\downarrow$ are spin-1/2, $\Uparrow/\Downarrow$ are spin-1 and bold arrows are the spin-3/2. 
In each manifold, the dimer-spins fluctuate back and forth and exchange with spins on both sides when it is present. 

The lowest energy state of (even, even) chain on the left part of Fig.~\ref{fS2}(a) has a moment that simply fluctuates back and forth while not contributing much to the magnetization. 
The first excited state $S^z=3/2$ of (even, even) chain on the right part of Fig.~\ref{fS2}(a) has a strong ferromagnetic correlation between $S_L$ and $S_R$. 
For the (odd, odd) chain the lowest energy state (left part of Fig.~\ref{fS2}(b)) carries spin-1/2 on both chains so that 
the ferromagnetic correlation develops mediated by the dimer-spin, 
which is further enhanced in the $S^z=3/2$ excited state(right part of Fig.~\ref{fS2}(b)). 
The (even, odd) case has a singlet lowest energy state where all the spins die out, 
but the excited state has a strong ferromagnetic correlation between all spins. 
Figure~\ref{fS2}(d) shows the magnetization $\langle S_L\rangle$ of the two lowest energy states, 
for different $N$ and different series of $N_L=N_R=$even, odd and $N_L+1=N_R$, which correspond to 
the above mentioned three cases. 
The correlation between these moments are mostly ferromagnetic, $\langle S_L\cdot S_R\rangle>0$, 
which is displayed in Fig.~5(c) in the main text for the same parameters by the same symbols. 
%

\subsection{Phenomenological treatment}

The lack of experimental information on the correlation length $\xi$ of charge ordering 
makes it difficult to precisely evaluate the thermodynamic quantities in theory. 
Here, we will make reasonable {\it assumption} that the correlation length grows rapidly 
on lowering the temperature as $\xi \propto (k_BT)^{-\nu}$, where $\nu=1$ is the critical exponent of 
the two-dimensional Ising universality class which the charge ordering transition belongs to. 
As we discussed in the introduction part of the main text, this second order phase transition is masked at 
low temperature in the real material. 
Therefore, although the form $\xi \propto (k_BT)^{-\nu}$ diverges with $T\rightarrow 0$, 
the increase of true $\xi$ will gradually slow down and should stop at some temperature. 
In the following, we make use of $\xi \propto (k_BT)^{-\nu}$, while supposing that 
this assumption may apply only at $T\gtrsim T^* \sim 0.15J_{\rm AF}$. 
Although the thermodynamic susceptibility we derive here in a phenomenological manner is fragile, 
we discuss it here because we find that it may help
the understanding of the possible behavior of $\chi$ that originates 
from the ferromagnetic correlation of the excited state of the model (1). 

\par
Let us start with the square-root form of the experimental $M$--$H$ curve, 
which agrees well with the energetics of our microscopic model; 
in Fig.5(e) in the main text, we showed the comparison between the experimental and theoretical data. 
From the latter data, we evaluate the functional form as 
$M/N = \alpha (\Delta E/N)^{1/2}$, with $\alpha=0.56$. 
Here, we denote the magnetization $M$ as a continuous variable, which can be identified as discrete 
variable $S_z$ used in the main text.

\par
The continuous $M$--$H$ curve gives the phenomenological shape of the $M$-dependence 
of the energy at zero field, $E(M,H=0)$. 
Since $H=\frac{\partial E(M,H=0)}{\partial M}= (M/\alpha N)^2$, 
we find $E(M,H=0)= N (M/N)^3 /(3\alpha^2)+{\rm const}$. 
From this argument, one can generate a series of discrete lowest energy levels for a given $M$-sector, 
$E(M,H=0)$, for various system size $N$, as shown in Fig.~\ref{fS3}(a). 
One finds that the envelope of the energy landscape is nearly flat at $M \lesssim 3$, particularly when $N$ is large, 
indicating that the system is indeed close to the partially polarized ferromagnetic long range ordered phase. 

\par
Let us regard the system as an ensemble of these chains. 
Although each spin chain is correlated with more than two other chains in reality, 
we use a rough assumption that the $N$-insensitive functional form of the two-chain calculation 
shown in Fig.~\ref{fS3}(a) will hold.
Within this assumption $\chi$ is determined by a typical length scale $N$ of the chains. 
$N$ can be read off as $\xi$ since the length of the antiferromagnetically coupled chain 
corresponds to the length scale of the charge ordering. 
We assume the gaussian distribution of $N$ as 
\begin{equation}
	P(N) \propto {\rm e}^{-(N-\xi)/(2\sigma^2)}, \hspace{5mm} 
	\xi= A/(k_BT)
	\label{prob}
\end{equation}
and by varying $A$ and $\sigma$, one can evaluate the partition function as 
$Z= \sum_N P(N) \sum_{M} {\rm e}^{-\beta E(M,H)}$, and the corresponding expectation value of 
the magnetization $\langle M \rangle$. 
By setting $H=0.004$ in unit of $J_{\rm AF}=1$ which corresponds to the experimental $H=1$\,T at which 
the susceptibility is measured, we obtain $\chi=\langle M \rangle/H$. 
Figure\,\ref{fS3}(b) shows the susceptibility $\chi$ obtained by varying $A=5$--10, 
where $k_BT \sim 0.5$ corresponds to about 90\,K when we interpret the energy unit as $J_{\rm AF}=170$\,K. 
One finds that $\chi^{-1}\propto (T-T_c)$ in the intermediate temperature region 
that resembles the experimental observation (Fig. 2(c) in the main text), 
while the actual value of $T_c$ depends on the parameter $A$ and also slightly on $\sigma$. 
The inverse of temperature dependence of $\bar N=\sum_{N=N_{\rm min}}^{N_{\rm max}} N P(N) \sim \xi$ 
we adopted for a given $A$ and $\sigma$ is shown in the inset of Fig.~\,\ref{fS3}(b). 
At the highest temperature $k_BT \sim 0.5$, the chain length is as short as $\xi \sim 10$--20 for $A=5$--10, 
and at $k_BT\sim 0.1$ it grows up to $\xi \sim 50$--100, where we took $(N_{\rm min},N_{\rm max})=(4,500)$. 
We mask the low temperature part in the figure since 
the assumption on the functional form of $\xi$ may at most hold only at 
$k_BT \gtrsim 0.15 J_{\rm AF}$, which is approximately the region $T\gtrsim T^*$, 
where $\bar N\lesssim 70$ remains short.

\end{document}